\documentclass[review]{elsarticle}
\usepackage{amsmath,amssymb}
\usepackage{lipsum,soul}
\usepackage{bm}
\usepackage{graphicx}
\usepackage{graphics}
\usepackage{amsmath}
\usepackage{array}
\usepackage[caption=false]{subfig}
\usepackage{xcolor}
\usepackage{subfig}
\usepackage{hyperref}
\usepackage[capitalise]{cleveref}
\usepackage{textcomp} % regestered trademark

\usepackage{epstopdf}
\usepackage{cases}
\usepackage{amssymb}
\usepackage{multirow}
\usepackage{siunitx}
\usepackage{cases}
\usepackage{tabularx}
\usepackage{nomencl}
\usepackage[top=1.1in, bottom=1.1in, left=1.0in, right=1.0in]{geometry}
\usepackage[utf8]{inputenc}
\usepackage{tikz}
\usepackage{comment}
\usepackage{lineno,hyperref}
\modulolinenumbers[5]
\usepackage{booktabs}

\newcommand{\lei}[1]{{\leavevmode\color{black}#1}}
\newcommand{\chuang}[1]{{\leavevmode\color{black}#1}}

\begin{document}

\title{A fast synthetic iterative scheme for the stationary phonon Boltzmann transport equation}
\author[add1]{Chuang Zhang}
\ead{zhangcmzt@hust.edu.cn}
\author[add1]{Songze Chen}
\ead{jacksongze@hust.edu.cn}
\author[add1]{Zhaoli Guo\corref{cor1}}
\ead{zlguo@hust.edu.cn}
\author[add2]{Lei Wu\corref{cor1}}
\ead{lei.wu.100@strath.ac.uk}
\cortext[cor1]{Corresponding author}
\address[add1]{State Key Laboratory of Coal Combustion, Huazhong University of Science and Technology,Wuhan, 430074, China}
\address[add2]{James Weir Fluids Laboratory, Department of Mechanical and Aerospace Engineering, University of Strathclyde, Glasgow G1 1XJ, UK}

\date{\today}

\begin{abstract}
	
\lei{The heat transfer \chuang{in solid materials} at the micro- and nano-scale can be described by the mesoscopic \chuang{phonon} Boltzmann transport equation (BTE), rather than the macroscopic Fourier's heat conduction equation that works only in the diffusive regime.} The implicit discrete ordinate method (DOM) is efficient \lei{to find the steady-state solutions of the BTE} for highly non-equilibrium heat transfer problems, but converges \lei{extremely} slowly in the near-diffusive regime.
In this paper, a fast synthetic iterative scheme is developed to accelerate convergence for the implicit DOM based on the stationary phonon BTE.
The key \lei{innovative point} of the present scheme is the introduction of the macroscopic \lei{synthetic diffusion} equation \lei{for the temperature, which is obtained from the zero- and first-order moment} equations of the phonon BTE. \lei{The synthetic diffusion equation, which is asymptomatically preserving to the Fourier's heat conduction equation in the diffusive regime, contains a term related to the Fourier's law and a term determined by the second-order moment of the distribution function that reflects the non-Fourier heat transfer. The mesoscopic kinetic equation and macroscopic diffusion equations are tightly coupled together, because the diffusion equation provides the temperature for the BTE, while the BTE provides the high-order moment to the diffusion equation to describe the non-Fourier heat transfer.}
This synthetic \lei{iterative scheme} strengthens the coupling of all phonons in the phase space \lei{to facilitate the fast convergence from the diffusive to ballistic regimes}.
Typical numerical tests \lei{in one-, two-, and three-dimensional problems demonstrate} that \lei{our} scheme can describe the multiscale heat transfer problems accurately and efficiently. For all test cases convergence is reached within one hundred iteration steps, which is one to three orders of magnitude faster than the traditional implicit DOM in the near-diffusive regime.

\end{abstract}
\maketitle
%\linenumbers

\section{INTRODUCTION}

The Boltzmann transport equation \lei{for} heat carriers, including photon, electron, neutron, phonon and so on, is widely used \lei{to model} multiscale energy transport and conversion~\cite{kaviany_2008,ChenG05Oxford,zhangZm07HeatTransfer,laurendeau2005statistical}.
In semiconductor devices, the phonon is regarded as the main heat carrier and the phonon Boltzmann transport equation (BTE)~\cite{Majumdar98MET,ZimanJM60phonons,srivastava1990physics} is usually used to predict the \chuang{multiscale} heat transfer in materials~\cite{cahill2003nanoscale,cahill2014nanoscale}, \lei{while the Fourier's heat conduction equation is only \chuang{valid} in the diffusive regime, i.e. when the system size is \chuang{much larger than the phonon mean free path}.}
For steady problems, the phonon BTE is composed of the \lei{advection and scattering terms} with six degrees of freedom, namely, the physical space ($x,~y,~z$ coordinates) and the wave vector space (frequency space and solid angle space)~\cite{MurthyJY05Review}.
Due to its complicated mathematical expression and multi-variables~\cite{narumanchi2005comparison,ChenG05Oxford}, it is important to numerically solve the phonon BTE efficiently and accurately for actual thermal applications.

\chuang{Many numerical methods, including the Monte Carlo method~\cite{MazumderS01MC,graphene_full_dispersion}, discrete ordinate method (DOM)~\cite{SyedAA14LargeScale}, discrete unified gas kinetic scheme (DUGKS)~\cite{GuoZl16DUGKS,LUO2017970}, and lattice Boltzmann method (LBM)~\cite{GuoYy16LB}, have been developed to solve the phonon BTE.
The Monte Carlo method is the most widely used one in micro/nano scale heat transfer because it can handle with the complex phonon dispersion and scattering physics easily and accurately. \lei{However, the requirement that
the time step and cell size have to be respectively smaller than the relaxation time and the phonon mean free path restricts its applications in the near-diffusive regime.
The Monte Carlo method suffers large statistics errors and converges very slowly in this regime. To fix this problem,}
some strategies~\cite{Lacroix05,PJP11MC,randrianalisoa2008monte} were proposed, such as the energy-based variance-reduced Monte Carlo method~\cite{PJP11MC,Hadjiconstantinou15MC}.
In this method, the stochastic particle description solves only the deviation from \lei{the equilibrium state} so that it \lei{reduces} the \lei{statistical error} significantly and converges \lei{faster than} the traditional Monte Carlo method when the temperature difference in the \lei{simulation} domain is small.
The \lei{DOM employs the deterministic discretization of the wave vector space and physical space, hence it is free of noise. However,  the discretization of the six-dimensional non-equilibrium distribution function needs lots of computer memory. Moreover, the phonon advection and scattering are handled separately so that it has large numerical dissipations in the near-diffusive regime, i.e. the numerical heat conductivity is much larger than the physical conductivity.}
To solve this problem, the DUGKS~\cite{GuoZl16DUGKS,LUO2017970}, \lei{which couples the phonon scattering and advection together at the cell interface within one time step, has been developed.}
It works well for all regimes and its time step is not restricted by the relaxation time. The lattice Boltzmann method~\cite{AdamC10MultiscaleLB,Ankur14nonequilibrium} works well in the near-diffusive regime but it is very hard to capture the multiscale phonon transport physics correctly with a wide range of group velocities and mean free paths, \lei{since its use of highly optimized but limited number of discrete solid angles.}

%
%Apart from above explicit methods, some other methods~\cite{Pareekshith16BallisticDiffusive,YangRg05BDE,Minnich15advances,AdamC10MultiscaleLB,Ankur14nonequilibrium} were developed to promote the development of numerical heat transfer, too. {\color{red}please expand with more details!}

For steady heat transfer problems, explicit methods~\cite{Pareekshith16BallisticDiffusive,Minnich15advances} usually converge slowly due to the limitation of the time step \lei{by the Courant–Friedrichs–Lewy condition}. The implicit iterative scheme, which has no such limitation, will be an excellent choice to find the steady state solution quickly. }
One of the most popular implicit methods is the implicit \lei{DOM}~\cite{Stamnes88,terris2009modeling,wangmr17callaway,FivelandVA96Acceleration}.
\lei{Given the initial temperature distribution}, the phonon BTE for each discretized wave vector is solved iteratively in the whole discretized physical space~\cite{MurthyJY12HybridFBTE}.
\lei{After each iteration},  the total energy or temperature is updated by the moment of the distribution function over the wave vector space based on the energy conservation of the phonon scattering term.
These processes are repeated till convergence.
This method converges very fast in the ballistic regime, since \lei{the phonon scattering is rare} and the information exchange in the physical space is efficient.
Note that the energy conservation of the scattering term is not satisfied numerically until \lei{the steady state is reached} when solving the phonon BTE iteratively for each discretized wave vector~\cite{MurthyJY15COMET,FivelandVA96Acceleration,EdwardW84synthetic,Chuang17gray}, which indicates that the coupling of the phonons with different wave vectors are inefficient.
Therefore, the iteration converges very slowly in the regimes \lei{where} the phonon scattering dominates the heat transfer, for example, in the near-diffusive regime.
\lei{In real materials}, unfortunately, the phonon mean free paths  \lei{span} several orders of magnitude~\cite{pop2004analytic,chung2004role}.
In other words, the phonon BTE is essentially multiscale and for actual thermal engineering, it is necessary to \lei{tackle the low efficiency problem in} the implicit DOM in the near-diffusive regime.

To accelerate convergence for the implicit DOM in the near-diffusive regime, many strategies~\cite{FivelandVA96Acceleration,ADAMS02fastiterative} have been developed.
One of them is the hybrid Fourier-BTE method~\cite{MurthyJY12HybridFBTE}, in which a cutoff Knudsen number is introduced and different equations are used to describe phonon behaviors with different mean free paths.
For phonon in each discretized wave vector, if the associated Knudsen number ($\text{Kn}$, the ratio of the mean free path to the characteristic length of the system) is larger than the cutoff one, the traditional implicit DOM is used; \lei{otherwise}, a modified Fourier equation is used to describe the thermal transport of phonons with \lei{small} mean free paths.
\lei{The hybrid Fourier-BTE} works well and accelerates convergence \lei{in both} the ballistic and diffusive regimes. \lei{However}, the choice of the cutoff Knudsen number \lei{that will affect the convergent solution has not been justified rigorously.}
%Another method is the coupled ordinate method (COMET)~\cite{MurthyJY15COMET}, which was first developed for radiation transport~\cite{MurthyJY99COMET}.
Different from the explicit numerical treatment of the scattering term in the implicit DOM, the coupled ordinate method (COMET)~\cite{MurthyJY15COMET}, which was first developed for radiation transport~\cite{MurthyJY99COMET}, \lei{employs the fully implicit treatment} on the scattering term in order to numerically ensure the energy conservation of the scattering term.
The relationships among the phonon distribution function, equilibrium state and the macroscopic variables are built over the whole wave vector space, and a huge coefficient matrix will be generated and solved iteratively.
This method realizes the efficient phonon coupling \lei{in both} the physical and wave vector spaces and accelerates convergence for all Knudsen numbers. \lei{However, it is not easy to solve} so many equations with different wave vectors simultaneously.

Another accelerate strategy for the implicit DOM is the synthetic method, which was first proposed for neutron transport~\cite{DSAneutron} and then developed for radiative heat
transfer applications~\cite{domradiation,DSAnuclear,coupledMS05} \lei{and rarefied} gas dynamics~\cite{DVMaccGas,WU2017431,WeiSu2019,ZhuYj16IUGKS}.
The main point of the synthetic scheme~\cite{ADAMS02fastiterative,EdwardW84synthetic,WU2017431,FivelandVA96Acceleration} is the introduction of the macroscopic \lei{moment} equations derived from the different \lei{moment equations of the BTE}, which strengthens the coupling of the heat carriers with different directions or frequencies~\cite{coupledMS05,ADAMS02fastiterative,FivelandVA96Acceleration}.
%One of most widely used synthetic method is the diffusion synthetic acceleration (DSA) method~\cite{EdwardW84synthetic}, which accelerates convergence rapidly for the implicit DOM, especially for problems with diffusion-like regimes.
Because the mathematical formulas of the phonon BTE under the relaxation time approximation is similar to the radiation transport equation with isotropic scattering \lei{and the Bhatnagar–Gross–Krook (BGK) kinetic model for gas dynamics~\cite{kaviany_2008,ChenG05Oxford}}, the synthetic idea will be a good start \lei{to find the steady-state solution of the phonon BTE}.
Recently, an implicit kinetic scheme, which is also a kind of synthetic scheme~\cite{DSAneutron}, was developed for multiscale heat transfer problems~\cite{Chuang17gray,ZHANG20191366}.
The zero-order moment equation of the phonon BTE, namely, the first-law of thermodynamics, is introduced to accelerate convergence for small Knudsen numbers.
Because no specific mathematical operator is used to represent the relationship between the heat flux and the temperature at the micro/nano scale, an approximate linear operator with artificial coefficient~\cite{RonSD82Newton} is constructed to diminish the macroscopic residual.
This method works for all Knudsen numbers and accelerates convergence in the near-diffusive regime compared to the implicit DOM.
However, the artificial coefficient has to be adjusted to ensure the convergence of the thermal transport in different regimes, as
the closer the approximate operator to the real operator, the faster the iteration converges~\cite{Andrew78Newton,RonSD82Newton},
%It is still an open question to find a better approximate linear operator to replace the real complex operator.

In this study, a fast synthetic iterative scheme is developed to accelerate convergence for the implicit DOM based on the stationary phonon BTE.
Motivated by the synthetic acceleration strategies~\cite{EdwardW84synthetic,ADAMS02fastiterative,DSAneutron,DVMaccGas,WU2017431,WeiSu2019}, the zero-order and first-order moment equations of the phonon BTE are \lei{combined to derive the diffusion equation for the temperature, which is asymptotically preserving to the Fourier's heat conduction equation in the diffusive regime. The macroscopic diffusion equation and the phonon BTE are solved \chuang{sequentially} to facilitate the fast convergence to the steady-state solutions.  The present scheme can capture the multiscale phonon transport accurately and efficiently, which is easy to implement as it requires few changes to the conventional implicit DOM.}

The rest of this article is organized as follows. In Sec.~\ref{scheme}, \lei{the phonon BTE, the synthetic iterative scheme,} and the boundary conditions are introduced and discussed in detail. In Sec.~3, the performances of the present scheme are tested by \lei{in a number of one-, two-, and three-dimensional multiscale heat transfer problems}. Finally, a conclusion is drawn in Sec.~\ref{conclusion}.

\section{NUMERICAL SCHEME}\label{scheme}

\subsection{Phonon Boltzmann transport equation}

For an isotropic \chuang{wave vector space}, the steady-state phonon BTE under the single-mode relaxation time approximation~\cite{peraud_monte_2014,MurthyJY05Review} is described by
\begin{equation}
v \bm{s} \cdot \nabla f= \frac{f^{eq}-f}{\tau},
\end{equation}
where $f\left( \bm{x}, \bm{s},\omega,p \right)$ is the phonon distribution function in the phase space, $\bm{x}$ is the spatial position, $\bm{s}$ is the unit direction vector, $v=\partial{\omega}/ \partial{k}$ is the group velocity, $k$ is the wave vector in one direction, $\omega$ is the angular frequency, $p$ is the phonon polarization, $\tau$ is the effective relaxation time, and $f^{eq}$ is the equilibrium \lei{distribution function given by the Bose-Einstein statistics}~\cite{ZimanJM60phonons,laurendeau2005statistical}.

For convenience we rewrite the BTE in energy density form,
\begin{equation}
v \bm{s} \cdot \nabla e= \frac{e^{eq}-e}{\tau},
\label{eq:eBTE}
\end{equation}
by introducing the energy distribution function $e=\hbar \omega D(\omega,p) \left[  f-f^{eq}(T_{\text{ref}})  \right]/ 4\pi$ and the associated equilibrium distribution function $e^{eq}=\hbar \omega D(\omega,p) \left[  f^{eq}-f^{eq}(T_{\text{ref}})  \right] / 4\pi$,
%as
%\begin{equation}
%e =\hbar \omega D(\omega,p) \left[  f-f^{eq}(T_{\text{ref}})  \right]/ 4\pi,
%\label{eq:ef}
%\end{equation}
%\begin{equation}
%e^{eq}=\hbar \omega D(\omega,p) \left[  f^{eq}-f^{eq}(T_{\text{ref}})  \right] / 4\pi,
%\label{eq:efeq}
%\end{equation}
where $\hbar$ is the Planck’s constant divided by $2\pi$, $T_{\text{ref}}$ is the reference temperature,  and $D(\omega,p)= {k^2}/(2\pi^{2}{v})$ is the phonon density of state.

Assuming the temperature difference in the domain is much smaller than the reference temperature \lei{$T_{\text{ref}}$} of the system, i.e., $\Delta T \ll T_{\text{ref}}$, then \lei{the relaxation time} $\tau \approx \tau(\omega, p, T_{\text{ref}})$ is approximately independent of the temperature, and
the equilibrium \lei{distribution function} can be linearized as
\begin{equation}
e^{eq} \approx C(\omega,p, T_{\text{ref}})\frac{T-T_{\text{ref}}} {4\pi},
\label{eq:linearized}
\end{equation}
where $C(\omega,p, T_{\text{ref}})=\hbar \omega D(\omega,p) {\partial{f^{eq}} }/ {\partial{T}}$ is the mode specific heat at $T_{\text{ref}}$ and $T$ is the temperature~\cite{MurthyJY05Review,ChenG05Oxford,peraud_monte_2014}. Due to the energy conservation of the scattering term, we have $\sum_{p}\int_{\omega_{min,p}}^{\omega_{max,p}} \int_{4\pi}  { (e^{eq}-e) }/{\tau} d{\Omega}d{\omega}=0$,
%\begin{equation}
%\sum_{p}\int_{\omega_{min,p}}^{\omega_{max,p}} \int_{4\pi} \frac { e^{eq}-e }{\tau} d{\Omega}d{\omega}=0,
%\label{eq:energyconser}
%\end{equation}
where $\omega_{min,p}$ and $\omega_{max,p}$ are the minimum and maximum frequency for a given phonon polarization branch $p$, \lei{respectively}, and $\Omega$ is the solid angle in spherical coordinates. Therefore, the temperature can be obtained by
\begin{equation}
T=T_{\text{ref}}+ \left( \sum_{p}\int_{\omega_{min,p}}^{\omega_{max,p}}  \frac{\int_{4\pi} e d\Omega }{\tau } d{\omega}   \right) \times \left(  \sum_{p}\int_{\omega_{min,p}}^{\omega_{max,p}}  \frac{ C}{\tau }  d{\omega} \right)^{-1}.
\label{eq:calT}
\end{equation}
The total heat flux is calculated by
\begin{equation}
\bm{q}= \sum_{p}\int_{\omega_{min,p}}^{\omega_{max,p}} \int_{4\pi} v \bm{s} e d{\Omega}d{\omega}.
\label{eq:heatflux}
\end{equation}
Note that from Eq.~\eqref{eq:eBTE}, we can obtain $\nabla \cdot \bm{q} =0$ due to the conservation of the scattering term.

\subsection{Phonon dispersion and scattering}

In this work, the phonon dispersion curves of the monocrystalline silicon in the [1 0 0] direction are chosen to represent the other directions~\cite{brockhouse1959lattice,chung2004role}.
Only the acoustic phonon branches, namely, longitude acoustic branch (LA) and transverse acoustic branch (TA), are considered and the dispersion curve reported in~\cite{pop2004analytic} is used, i.e.,
\begin{equation}
\omega=c_{1}k+c_{2}k^2,
\label{eq:curves}
\end{equation}
where $k \in [0, 2\pi/a]$, $a=0.543\text{nm}$, $c_{1}$ and $c_{2}$ are two coefficients.
For LA, $c_{1}=9.01 \times 10^5$cm/s, $c_{2}=-2.0 \times 10^{-3}$$\text{cm}^{2}$/s; for TA, $c_{1}=5.23 \times 10^5$cm/s, $c_{2}=-2.26 \times 10^{-3}$$\text{cm}^{2}$/s~\cite{pop2004analytic}.
Apart from above parameters, the phonon scattering~\cite{holland1963analysis} is important for the solution of the phonon BTE, too.
The Matthiessen's rule is used to couple all phonon scattering mechanisms together~\cite{MurthyJY05Review}, i.e.,
\begin{equation}
\tau^{-1}=\tau_{{\text{impurity}}}^{-1}+\tau_{{\text{U}}}^{-1}+\tau_{{\text{N}}}^{-1},
\label{eq:Matthiessen}
\end{equation}
where the specific formulas of impurity scattering $\tau_{{\text{impurity}}}$, U scattering $\tau_{{\text{U}}}$ and N scattering $\tau_{{\text{N}}}$ can refer to Ref~\cite{terris2009modeling}.

\subsection{Implicit discrete ordinate method}

The stationary phonon BTE~\eqref{eq:eBTE} is usually solved by the implicit DOM~\cite{Stamnes88,wangmr17callaway,terris2009modeling}, in which the frequency space and the solid angle space are discretized into lots of small pieces with certain quadrature rules, respectively.
%The phonon dispersion is also important for the heat transfer in real materials~\cite{chung2004role}.
For each phonon branch $p$, the wave vector $k$ is discretized equally into $N_B$ discrete bands, i.e., $k_b=2\pi(2b-1)/(2a N_B)$, where $b \in [1,N_B]$.
Based on Eq.~\eqref{eq:curves}, we can obtain $\omega_{b}=c_{1}k_b+c_{2}k_{b}^2$, $v_{\omega_b,p}=c_{1}+2c_{2}k_{b}$ and $\tau^{-1}=\tau^{-1}_{\omega_b,p}(T_{\text{ref}})$.
The mid-point rule is used for the numerical integration of the frequency space.
For the solid angle space in spherical coordinates, we set $\bm{s}=\left( \cos \theta, \sin \theta \cos \varphi, \sin \theta  \sin \varphi  \right)$, where $\theta \in [0,\pi]$ is the polar angle and $\varphi \in [0,2\pi]$ is the azimuthal angle.
The $\cos \theta \in [-1,1]$ is discretized with the $N_{\theta}$-point Gauss-Legendre quadrature~\cite{Abramovitch65Math,NicholasH13GaussL}, while the azimuthal angular space $\varphi \in [0,\pi]$ (due to symmetry) is discretized with the $\frac{N_{\varphi}}{2}$-point Gauss-Legendre quadrature.
%The total discretized directions are $N_{dir}=N_{\theta} \times N_{\varphi}$.
Then we have $N_{dir}=N_{\theta} \times N_{\varphi}$ discretized directions $\bm{s}_{\alpha}$, where $\alpha \in [1,N_{dir}]$.

Giving a macroscopic temperature $T^{n}$ at the $n$-th iteration step, the distribution function at the next iteration step $e^{n+1}$ for a given discretized frequency band and direction is updated by solving the following equation
\begin{equation}
e^{n+1}_{\alpha,\omega_b,p}+{\tau}_{\omega_b,p} v_{\omega_b,p} \bm{s}_{\alpha} \cdot \nabla e^{n+1}_{\alpha,\omega_b,p}= e^{eq}_{\omega_b,p}(T^n).
\label{eq:SIBTE}
\end{equation}
\lei{We apply the following finite volume scheme to discretize Eq.~\eqref{eq:SIBTE}}:
\begin{equation}
e^{n+1}_{i,\alpha,\omega_b,p}+{\tau}_{\omega_b,p} v_{\omega_b,p}  \frac{1}{V_i} \sum_{j\in N(i)} S_{ij} \mathbf{n}_{ij}  \cdot  \bm{s}_{\alpha} e^{n+1}_{ij,\alpha,\omega_b,p} = e^{eq}_{i,\omega_b,p}(T^n),
\label{eq:DiSIBTE}
\end{equation}
where $V_i$ is the volume of cell $i$, $N(i)$ denotes the sets of face neighbor cells of cell $i$, $ij$ denotes the interface between cell $i$ and cell $j$, $S_{ij}$ is the area of the  interface $ij$, $\mathbf{n}_{ij}$ is the normal of the interface $ij$ directing from cell $i$ to cell $j$.
The van Leer limiter~\cite{Numericalanalysis} is used to calculate the distribution function at the cell interface $e_{ij}$ for numerical accuracy and stability.
The detailed solution of Eq.~\eqref{eq:DiSIBTE} can refer to some previous references~\cite{YoonS88LUSGS,MurthyJY98FVMradiative,Stamnes88}.

Based on Eq.~\eqref{eq:calT}, the temperature $T^{*}$ can be obtained by
\begin{equation}
T^{*}=T_{\text{ref}}+ \left( \sum_{p} \sum_{b=1}^{N_B} w_b \frac{ \sum_{\alpha=1}^{N_{dir}} e^{n+1}_{\alpha,\omega_b,p} w_{\alpha}  }{ \tau_{\omega_b,p} }    \right) \times \left(  \sum_{p} \sum_{b=1}^{N_B} w_b  \frac{C_{\omega_b,p}}{\tau_{\omega_b,p} }  \right)^{-1},
\label{eq:disT}
\end{equation}
where $w_b$ and $w_{\alpha}$ are the associated weights of discretized frequency space and solid angle space, respectively.
%\lei{is it better to introduce this in the first paragraph of Sec. 2.2, when you introduce the discretized solid angle and frequency???}
The heat flux is updated by
\begin{equation}
\bm{q}^{*}= \sum_{p} \sum_{b=1}^{N_B} w_b v_{\omega_b,p}  \sum_{\alpha}^{N_{dir}} w_{\alpha}  \bm{s}_{\alpha} e^{n+1}_{\alpha,\omega_b,p}.
\label{eq:disq}
\end{equation}
{\chuang{In the implicit DOM, we set $T^{n+1}=T^{*}$, $\bm{q}^{n+1}=\bm{q}^{*}$.
The above process is repeated till convergence.}}

The implicit DOM converges very fast for the heat transfer in the ballistic regime, \lei{however,} the number of the iteration steps increases significantly as the \lei{system size is much larger than the mean free path of phonons}~\cite{EdwardW84synthetic,WU2017431,FivelandVA96Acceleration,MurthyJY15COMET,MurthyJY12HybridFBTE}.
\lei{Our goal in the present work is} to develop a fast iterative scheme to accelerate convergence for the implicit DOM in the near-diffusive regime.

\subsection{Synthetic diffusion equation for the temperature}

\lei{One of the reasons for the slow convergence of the traditional iterative scheme~\eqref{eq:SIBTE} is that, at the $(n+1)$-th iteration step, the temperature is evaluated at the $n$-th step. To tackle this problem, a macroscopic diffusion equation for the temperature should be established; this equation should derived exactly from the mesoscopic phonon BTE, meanwhile, it should be to recover the Fourier's heat transfer law in the diffusive limit.}

\lei{To do this, let us recall that, when the phonon mean free path is much smaller than the characteristic system size, the Fourier's law is approximated obtained from the first-order Chapman-Enskog expansion~\cite{GuoZl16DUGKS,LUO2017970}, where the heat flux is
\begin{equation}
\bm{q} \approx\bm{q}_{\text{Fourier}}=-k_{\text{bulk}} \nabla T,
\end{equation}
where
\begin{equation}
k_{\text{bulk}}=\frac{1}{3} \sum_{p} \int_{\omega_{min,p}}^{\omega_{max,p}} Cv^{2}\tau d{\omega}
\label{eq:bulkconductivity}
\end{equation}
is the bulk thermal conductivity obtained in the diffusive limit. Note that $k_{\text{bulk}}$ is a constant under the assumption of $\Delta T \ll T_{\text{ref}}$. }

{\color{black} When the phonon mean free path is comparable to or even larger than  the characteristic system size, high-order contribution to the heat flux emerges, and the heat flux can be separated into the Fourier part and the non-Fourier part:
	\begin{equation}
	\begin{array}{l}
	\begin{split}
	\bm{q} =\bm{q}_{\text{Fourier}}+ \bm{q}_{\text{non-Fourier}}  =-k_{\text{bulk}} \nabla T + \bm{q}_{\text{non-Fourier}}.
	\label{eq:qsep}
	\end{split}
	\end{array}
	\end{equation}
}

\lei{The key to developing the synthetic diffusion equation is to find the expression for the non-Fourier part of heat flux, so that the diffusion equation for the temperature can be obtained by applying $\nabla \cdot \bm{q}=0$ to Eq.~\eqref{eq:qsep}, an equation which is exacly the zero-order \lei{moment equation} of Eq.~\eqref{eq:eBTE}.}

{\color{black}To express the heat flux in the form of Eq.~\eqref{eq:qsep}, the phonon BTE~\eqref{eq:eBTE} is firstly  multiplied by $\tau{v}\bm{s}$ and then integrated over the whole wave vector space, which leads to
\begin{align}
\sum_{p}\int_{\omega_{min,p}}^{\omega_{max,p}}  \int_{4\pi} \tau v^2  \bm{s}\bm{s}  \cdot \nabla e d\Omega  d\omega &= - \bm{q},  \label{eq:firstmomentum}  \\
 \nabla \cdot  \left( \sum_{p}\int_{\omega_{min,p}}^{\omega_{max,p}}  \int_{4\pi} \tau v^2  \bm{s}\bm{s}  \cdot \nabla e d\Omega  d\omega  \right) &= 0.
\end{align}
We reformulate Eq.~\eqref{eq:firstmomentum} as
\begin{equation}
\sum_{p}\int_{\omega_{min,p}}^{\omega_{max,p}}  \int_{4\pi}  \left[ \tau v^2  \bm{s}\bm{s} -A(\omega,p)\bm{I} \right] \cdot \nabla e  +  A(\omega,p) \bm{I} \cdot \nabla e  d\Omega  d\omega = - \bm{q},
\label{eq:macrokey}
\end{equation}
where $\bm{I}$ is the second order tensor of the unit, while the coefficients are chosen to be
\begin{equation}
A(\omega,p)= k_{\text{bulk}} \left( \tau  \sum_{p}\int_{\omega_{min,p}}^{\omega_{max,p}}  \frac{ C}{\tau }  d{\omega} \right)^{-1},
\label{eq:Awp}
\end{equation}
so that according to Eq.~\eqref{eq:calT} the last term on the left-hand side of Eq.~\eqref{eq:macrokey} is exactly the Fourier's law:
\begin{equation}
\sum_{p}\int_{\omega_{min,p}}^{\omega_{max,p}}  \int_{4\pi}  A(\omega,p)\tau \cdot \nabla \frac{e}{\tau}  d\Omega  d\omega =k_{\text{bulk}} \nabla T= - \bm{q}_{\text{Fourier}}.
\label{eq:qmacroA}
\end{equation}}

\lei{Clearly, the non-Fourier part of the heat flux is
\begin{equation}
 \bm{q}_{\text{non-Fourier}}=-\sum_{p}\int_{\omega_{min,p}}^{\omega_{max,p}}  \int_{4\pi}  \left[ \tau v^2  \bm{s}\bm{s} -A(\omega,p) \bm{I} \right] \cdot \nabla e  d\Omega  d\omega,
\label{eq:HoT}
\end{equation}
and the diffusion equation for the temperature is
\begin{align}
k_{\text{bulk}} \nabla^2 T  &= \nabla \cdot \left(  \bm{q}_{\text{non-Fourier}} \right).
\label{eq:governingE}
\end{align}

}

To sum up, we build the correct relationships among these macroscopic variables, i.e., $\bm{q}_{\text{non-Fourier}}$, $\bm{q}$ and $T$, by introducing the zero-order and first-order \lei{moment} equations of the phonon BTE.
Equation~\eqref{eq:governingE} indicates that the temperature can be calculated by the non-Fourier heat flux.
Although the real mathematical formula of the non-Fourier heat flux is unknown, it can be obtained by taking the \lei{moment} of the distribution function in the framework of phonon BTE \lei{according to} Eq.~\eqref{eq:HoT}.
In the diffusive limit, $\bm{q}_{\text{non-Fourier}}=0$ and the diffusion equation~\eqref{eq:governingE} recovers the traditional Fourier's heat conduction equation correctly.

Next, we will discuss the details of the solution of the macroscopic equation.
The finite volume method is used again to solve Eq.~\eqref{eq:governingE}, i.e.,
\begin{equation}
\sum_{j\in N(i)} S_{ij} \mathbf{n}_{ij}  \cdot  \nabla T_{ij}^{n+1}
= \frac{1}{k_{\text{bulk}} }  \sum_{j\in N(i)} S_{ij}  \mathbf{n}_{ij}  \cdot  \left( \bm{q}_{\text{non-Fourier}} \right)_{ij}^{*},
\label{eq:dvgoverningE}
\end{equation}
where $\bm{q}_{\text{non-Fourier}}$ is calculated by the second-order moment of the distribution function, i.e.,
\begin{equation}
\begin{array}{l}
\begin{split}
\left( \bm{q}_{\text{non-Fourier}} \right)_{ij}^{*} &= - \sum_{p}\int_{\omega_{min,p}}^{\omega_{max,p}}  \int_{4\pi}     \left[ \tau v^2  \bm{s}\bm{s} -A(\omega,p) \bm{I} \right] \cdot \nabla e_{ij}^{n+1}   d\Omega  d\omega   \\
&= - \sum_{p}  \sum_{b=1}^{N_B} w_b \sum_{\alpha=1}^{N_{dir}} w_{\alpha}    \left[ \tau_{\omega_b,p} v_{\omega_b,p}^2  \bm{s}_{\alpha}\bm{s}_{\alpha} -A_{\omega_b,p} \bm{I} \right] \cdot \nabla  e_{ij,\alpha,\omega_b,p}^{n+1}.
\label{eq:HoTface}
\end{split}
\end{array}
\end{equation}
The conjugate gradient method~\cite{datta2010numerical,ADAMS02fastiterative,Numericalanalysis} is used to solve the above equation for the update of the temperature and $10$ orders of magnitude reduction of residual are enforced.

\subsection{Boundary conditions}

\lei{The boundary condition plays an important role in the  heat transfer. Usually, the thermalization boundary condition, specular/diffusely reflecting boundary condition and the periodic boundary condition are considered in the phonon transport~\cite{GuoZl16DUGKS}}.
For the distribution function in Eq.~\eqref{eq:DiSIBTE}, detailed treatments of boundary conditions are the same as that in the traditional DOM~\cite{terris2009modeling,hsieh2012thermal}.

\begin{figure}
 \centering
 \includegraphics[scale=0.35,viewport=0 150 850 450,clip=true]{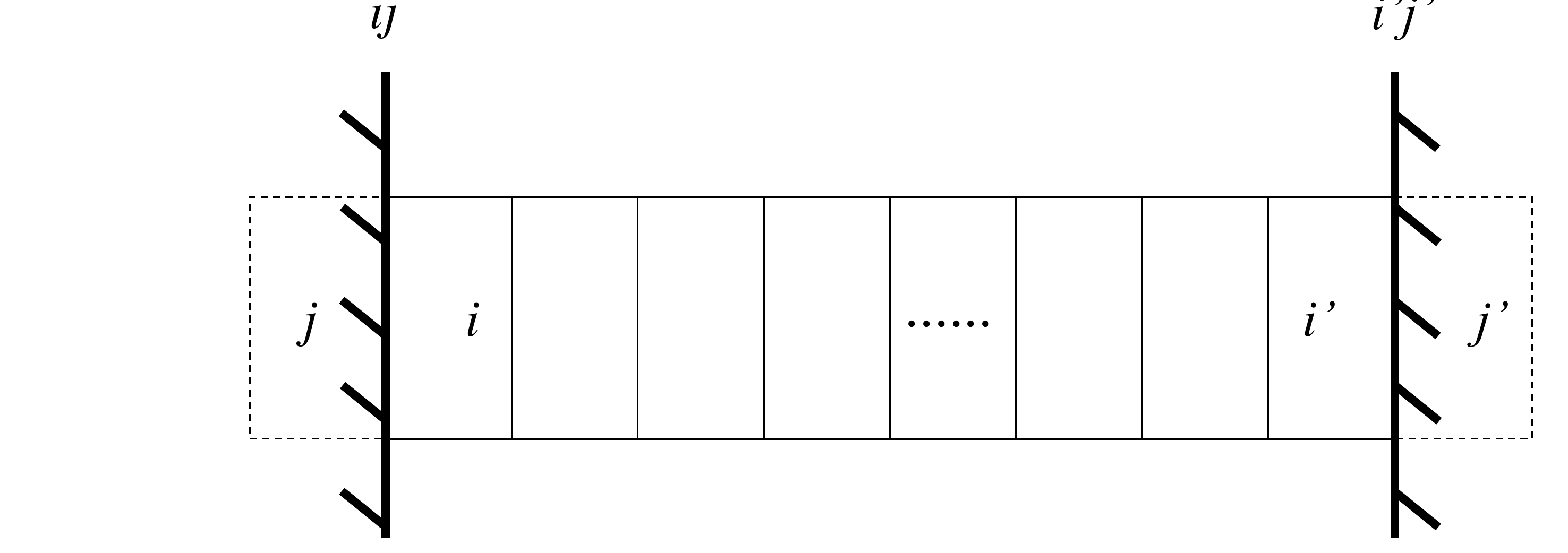}
 \caption{Ghost cells for boundary conditions.}
 \label{mesh_ghost}
\end{figure}

Here we focus on the boundary treatments of the macroscopic iteration, i.e., the solution of Eq.~\eqref{eq:dvgoverningE}.
Considering a boundary interface $ij$ between the ghost cell $j$ and the inner cell $i$ in \cref{mesh_ghost}, numerical treatments of different boundary conditions are presented as follows:
\begin{enumerate}
  \item The thermalization boundary is a kind of Dirichlet boundary condition with a fixed wall temperature $T_{\text{w}}$. However, in the non-diffusive regime, there is temperature jump on the boundary, i.e., $T_{ij} \neq  T_{\text{w}}$. Based on the moment of the distribution function $e_{ij}$, i.e., Eq.~\eqref{eq:disT}, we can calculate the temperature $T_{ij}$, then set $T_j=2 T_{{ij}}-T_i$.
%   \item  \chuang{ The specular/diffusely reflecting boundary condition belongs to the adiabatic boundary condition, which allows that the net heat flux across the boundary is zero, i.e., $$   \mathbf{n}_{ij} \cdot \bm{q}_{ij} =0.$$ Usually, we set $T_j=T_i$ and $ \mathbf{n}_{ij} \cdot \left( \text{HoT} \right) _{ij}=0$.   }

  \item \chuang{ The specular/diffusely reflecting boundary condition belongs to the adiabatic boundary condition, which \lei{requires} that the net heat flux across the boundary is zero,} i.e., $\mathbf{n}_{ij} \cdot \bm{q}_{ij} =0$. Thus we have $\mathbf{n}_{ij} \cdot \left(  k_{\text{bulk}} \nabla T_{ij}  \right) = \mathbf{n}_{ij} \cdot \bm{q}_{\text{non-Fourier},ij}$, and hence
      \begin{equation}
    \mathbf{n}_{ij} \cdot \left(  k_{\text{bulk}} \frac{T_j-T_i}{\bm{x}_j-\bm{x}_i }  \right) = \mathbf{n}_{ij} \cdot \bm{q}_{\text{non-Fourier},ij} .
      \end{equation}

  \item The periodic boundary condition usually involves two corresponding boundary interfaces, for example boundary interface $ij$ and its associated interface $i'j'$ between the ghost cell $j'$ and the inner cell $i'$, as shown in~\cref{mesh_ghost}. Two constraints can be derived: $T_{j'}-T_{i'j'}=T_i - T_{ij}$ and $T_j - T_{ij}=T_{i'}-T_{i'j'}$. If there is no temperature difference between the periodic boundaries, i.e., $T_{ij}=T_{i'j'}$, then we have $T_{j'}=T_{i}$, $T_j=T_{i'}$.
\end{enumerate}

\subsection{Solution procedure}

In summary, the main procedure of the present synthetic iterative scheme \lei{is} depicted as follows:
\begin{enumerate}
	\item give a reasonable macroscopic distribution, i.e., $T^{n}$;
	\item update the distribution function at the next iteration step $e^{n+1}$ based on Eq.~\eqref{eq:DiSIBTE};
	\item calculate $\bm{q}_{\text{non-Fourier}}$ based on Eq.~\eqref{eq:HoTface}, and update the temperature $T^{*}$ and the heat flux $\bm{q}^{*}$ based on Eqs.~\eqref{eq:disT} and~\eqref{eq:disq};
	\item update the temperature at the next iteration step $T^{n+1}$ based on Eq.~\eqref{eq:dvgoverningE};
	\item if converged, stop the iteration; otherwise, \lei{repeat step 2 to step 5}.
	%\item repeat step 2 to step 6.
\end{enumerate}
{\chuang{At the end of each iteration step, the heat flux $\bm{q}^{*}$ and the temperature $T^{*}$ are regarded as our finial results.}

{\chuang{In the present synthetic scheme, the macroscopic diffusion equation is introduced to accelerate convergence in the near-diffusive regime and coupled tightly with the phonon BTE.
The diffusion equation provides the temperature for the phonon BTE, while the phonon BTE provides the second-order moment to the diffusion equation to describe the non-Fourier heat transfer.
In the diffusion equation~\eqref{eq:dvgoverningE}, the Fourier part heat flux with temperature diffusion and the non-Fourier part heat flux are separated and calculated at two different iteration steps.
%The non-Fourier part heat flux dominates the thermal transport in the ballistic regime, while in the near-diffusive regime, the Fourier-part heat flux dominates the heat transfer.
%In the diffusive limit, the diffusion equation recovers the Fourier law correctly.
In the near-diffusive regime, the Fourier-part heat flux dominates the heat transfer and any disturbance of the temperature at one point can be quickly diffused by all other spatial points.
While in the ballistic regime, non-Fourier part heat flux dominates the thermal transport and the exchange of information through temperature diffusion \lei{is negligible}.
The combination of the diffusion equation and the implicit DOM makes the present scheme efficient for all regimes.
}}

\section{NUMERICAL TESTS}

In this section, we present some numerical simulations to \lei{assess the accuracy and efficiency} of the present scheme for multiscale heat transfer problems.
\chuang{The heat transfer in a 3D cube silicon material with side length $L$ is simulated.
In the $x$, $y$ and $z$ direction, there are left (right), top (bottom), front (back) boundary faces, respectively.}
The cartesian grids are used to discrete the physical space and $N_x$, $N_y$ and $N_z$ uniform cells are used for $x,~y,~z$ direction, respectively.
A parameter is introduced to measure the convergence
\begin{equation}
\epsilon=\frac{ \sqrt {\sum_{i}^{N_{cell}}{(T_i^{n}-T_i^{n+1})^2}}  } { \sqrt {   \sum_{i}^{N_{cell}}(\Delta T \times \Delta T)} },
\label{eq:epsilon}
\end{equation}
where $N_{cell}=N_x \times N_y  \times N_z$.
We assume as $\epsilon < 10^{-8}$ the system is converged.
Without special statements, in the following simulations we set $T_{\text{ref}}=300\text{K}$ and the initial temperature distribution in the domain is $T_{\text{ref}}$.
\lei{At} this temperature, the phonon mean free paths of silicon in different frequencies range from tens of nanometers to hundreds of microns.
As the characteristic length of the system is a few microns to a few tens of microns, the heat transfer is regarded as in the near-diffusive regime.
As $N_B \geq 20$, the numerical integration in the frequency space is regarded as converged based on the \lei{calculation of the} bulk thermal conductivity, i.e., Eq.~\eqref{eq:bulkconductivity}.
In our simulations, $k_{\text{bulk}} \approx 145.8$W/(m.K).
MPI paralleling computation with 24 cores (\chuang{Intel(R) Xeon(R) CPU E5-2680 v3 @ 2.50GHz}) based on the solid angle space is implemented and the CPU time mentioned in the following is the actual wall time for computation.
%\lei{More detailed description is needed!}

\subsection{One-dimensional case}

\begin{figure}[t]
     \centering
     \includegraphics[width=0.5\textwidth]{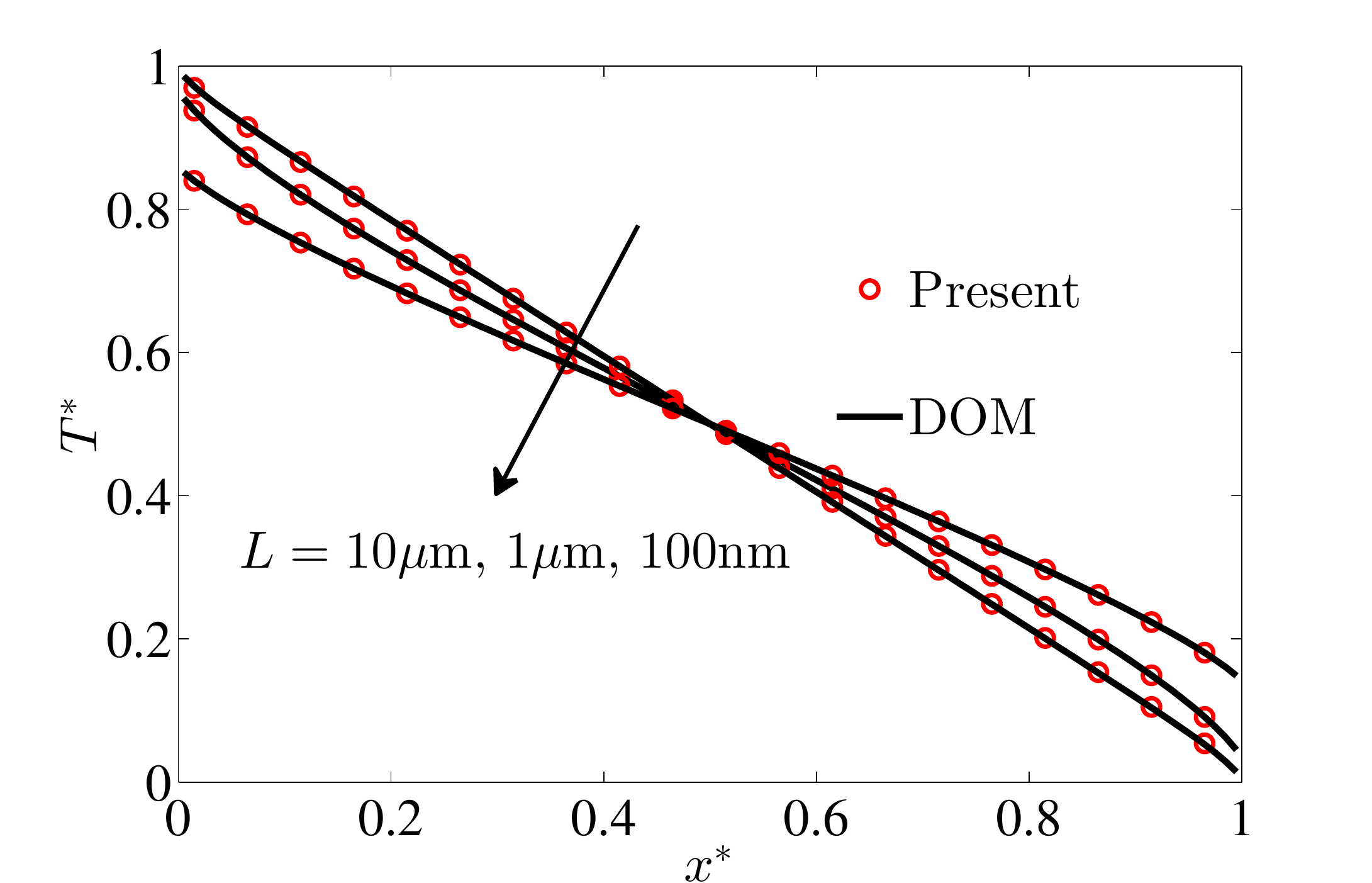}
     \caption{\lei{Temperature distributions in the quasi-one-dimensional cross-plane heat transfer with different length $L$, where $x^{*}=x/L$ and  $T^{*}=(T-T_R)/\Delta T$. For clarity, results from the present synthetic iterative scheme are shown at every 5 spatial cells.}  }
     \label{film1dnongray}
\end{figure}

The \lei{quasi-one-dimensional} cross-plane heat transfer is tested.
A temperature difference $\Delta T$ is implemented on the $x$ direction and the temperature of the left and right boundaries are set to be $T_L=T_{\text{ref}}+ \Delta T/2$ and $T_R=T_{\text{ref}}- \Delta T/2$, respectively.
Thermalization boundary conditions are used for these two boundaries. The other four boundaries are set to be periodic.
Then heat will transfer across the geometry from the left to the right.
In order to describe the thermal conduction process in different regimes, we set $N_x=100,~N_z=N_y=1$ and enough discretized directions are used with $N_{\theta}=40$ and $N_{\varphi}=8$. The phonon dispersion is included with $N_B=40$, which can capture the multiscale phonon transport physics correctly.

The numerical results are compared with the solutions of the implicit DOM in~\cref{film1dnongray}.
It can be observed that the temperature fields predicted by two methods match well with each other \lei{at typical length scales}.
In addition, the efficiency of the present scheme and the implicit DOM is also compared in different length scale, as summarized in Table.~\ref{1defficiency}.
It can be found that, compared to the implicit DOM, the present scheme has no acceleration in the ballistic regime.
Although the CPU time cost per iteration step by the present scheme increases $15-20$ percents due to the introduction of the macroscopic iteration, the present scheme accelerates convergence by one to three orders of magnitude in the transition and near-diffusive regimes.
As $L\geq 10 \mu$m, it is very \lei{difficult} for the implicit DOM to reach convergence.
However, for the present scheme convergence is reached within 100 iteration steps for all regimes.
%\lei{what is the convergence criterion of Eq.~\eqref{eq:HoTface}? In each iteration the iteration step for Eq.~\eqref{eq:HoTface} is different, is the time per step the average time per step??} }\vskip 0.2cm
\begin{table}
\caption{The efficiency of the present scheme in cross-plane heat transfer. Accelerate rate is the ratio of the total CPU time between the implicit DOM and the present scheme. Steps mean the total iteration number. \chuang{Time per step is the average CPU time cost for each iteration step}. }
\centering
\begin{tabular}{|*{8}{c|}}
 \hline
 \multirow{2}{*}{{\shortstack{$L$ }}}  & \multicolumn{3}{c|}{Present} & \multicolumn{3}{c|} {DOM} & \multirow{2}{*}{{\shortstack{Accelerate rate}}}\\
\cline{2-7}
 & Time (s) & Steps & Time per step   & Time (s) & Steps & Time per step  &  \\
 \hline
100 $\mu$m & 1.24 & 39 & 0.0318   & $>$2800  & $>$100000 & 0.028 & $>$ 2258    \\
 \hline
 10 $\mu$m & 2.26 & 71 & 0.0318    & 329.7 & 12164 & 0.0271   & 145.9  \\
  \hline
 5 $\mu$m & 2.30 & 73 & 0.0315    & 112.3 & 4108 & 0.0273   & 48.8  \\
\cline{1-3}
\cline{5-6}
\cline{8-8}
 \hline
1 $\mu$m & 2.14 & 67 & 0.0319 & 13.3 & 489 & 0.0272 & 6.2 \\
\cline{1-3}
\cline{5-6}
\cline{8-8}
 \hline
  500 nm & 2.03 & 63 & 0.0322 & 6.06 & 220 &  0.0275 & 3 \\
 \hline
 100 nm & 1.70 & 53 & 0.0321 & 1.58 & 56 & 0.0282 & 0.92 \\
 \hline
\end{tabular}
\label{1defficiency}
\end{table}

\subsection{Two-dimensional cases}

\subsubsection{In-plane heat transfer}

In-plane heat transfer is widely simulated in previous works.
A constant and small temperature gradient $\Delta T/L$ is \lei{applied in the $x$ direction, the temperature of the left and right boundaries are set to be $T_L=T_{\text{ref}}+ \Delta T/2$ and $T_R=T_{\text{ref}}- \Delta T/2$, respectively.}
The top and bottom faces are adiabatic and the others are periodic.
The diffusely reflecting boundary conditions are implemented on the adiabatic boundaries.
For the spatial space, we set $N_x=20,~N_y=100$ and $N_z=1$, which is enough to capture the multiscale heat transfer accurately.
The phonon dispersion is included with $N_B=40$ and $N_{\theta} \times N_{\varphi}=24 \times 24$ discretized directions are used.

The heat flux predicted by the present scheme is compared with the analytical solutions given in Refs.~\cite{Cuffe15conductivity,LUO2017970}.
Numerical results are shown in~\cref{heatfluxinplane} and excellent agreements can be observed \lei{at} different length scales.
In addition, the convergence history of the present scheme and implicit DOM is also compared in~\cref{historyinplane}.
As $L =100 \mu$m, it is not economic to used the implicit DOM, but the present scheme converges very fast.
As $L=10\mu$m, the CPU time cost by the implicit DOM is $10404$ seconds, while CPU time cost of the present scheme is only $67$ seconds, which is $155$ times faster than the former.
Besides, it can be found that the acceleration rate of the present scheme decreases with the decreasing of $L$, which indicates that the macroscopic diffusion equation loses its function as the Knudsen number increases.
In a word, it can be observed that in the near-diffusive regime, the convergence can be reached within $100$ steps.

\begin{figure}[t]
 \centering
 \subfloat[]{\includegraphics[width=0.33\textwidth]{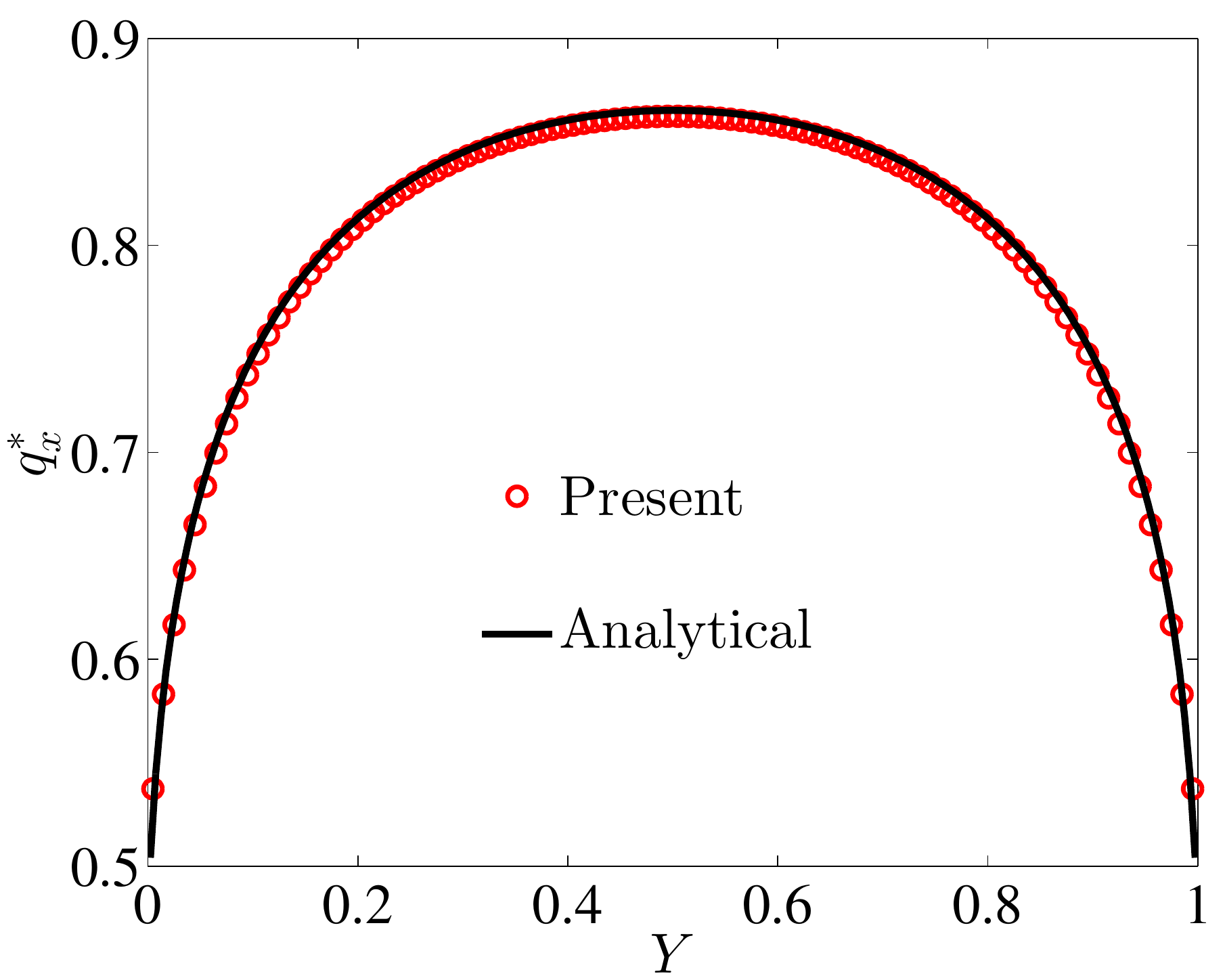}}~~
 \subfloat[]{\includegraphics[width=0.33\textwidth]{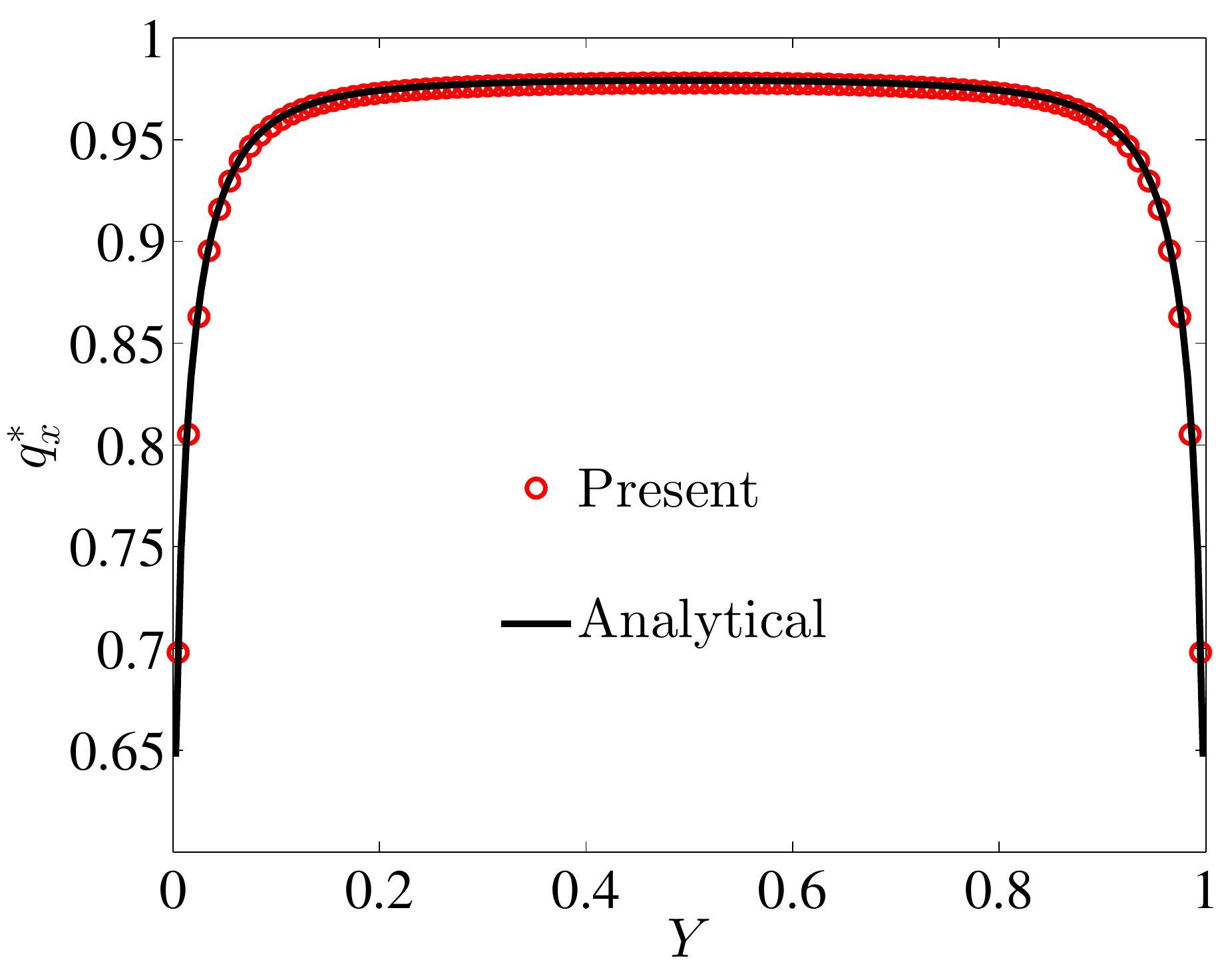}}~~
 \subfloat[]{\includegraphics[width=0.33\textwidth]{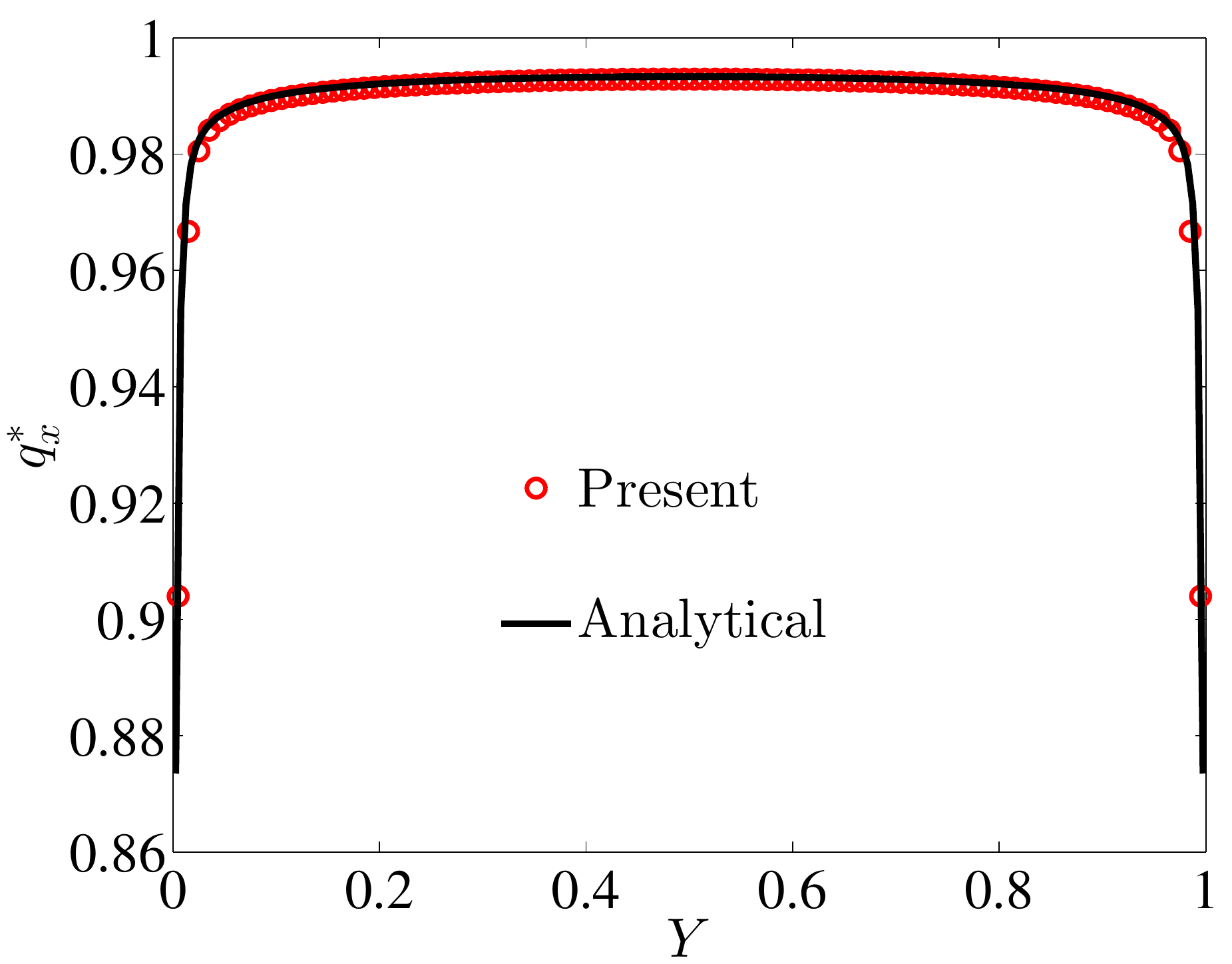}}~~
 \caption{The distribution of in-plane directional heat flux in the $y$ direction. $Y=y/L$, $q_{x}^{*}=q_{x}(Y)/q_{\text{bulk}}$, where $q_{\text{bulk}}= k_{\text{bulk}}\times \Delta T/L $. Red circle is the present numerical results and the black solid line is the analytical solution. (a)  $L= 1.0\mu$m, (b) $L=10\mu$m, (c) $L= 100\mu$m. }
 \label{heatfluxinplane}
\end{figure}

\begin{figure}[t]
     \centering
     \subfloat[]{\includegraphics[width=0.45\textwidth]{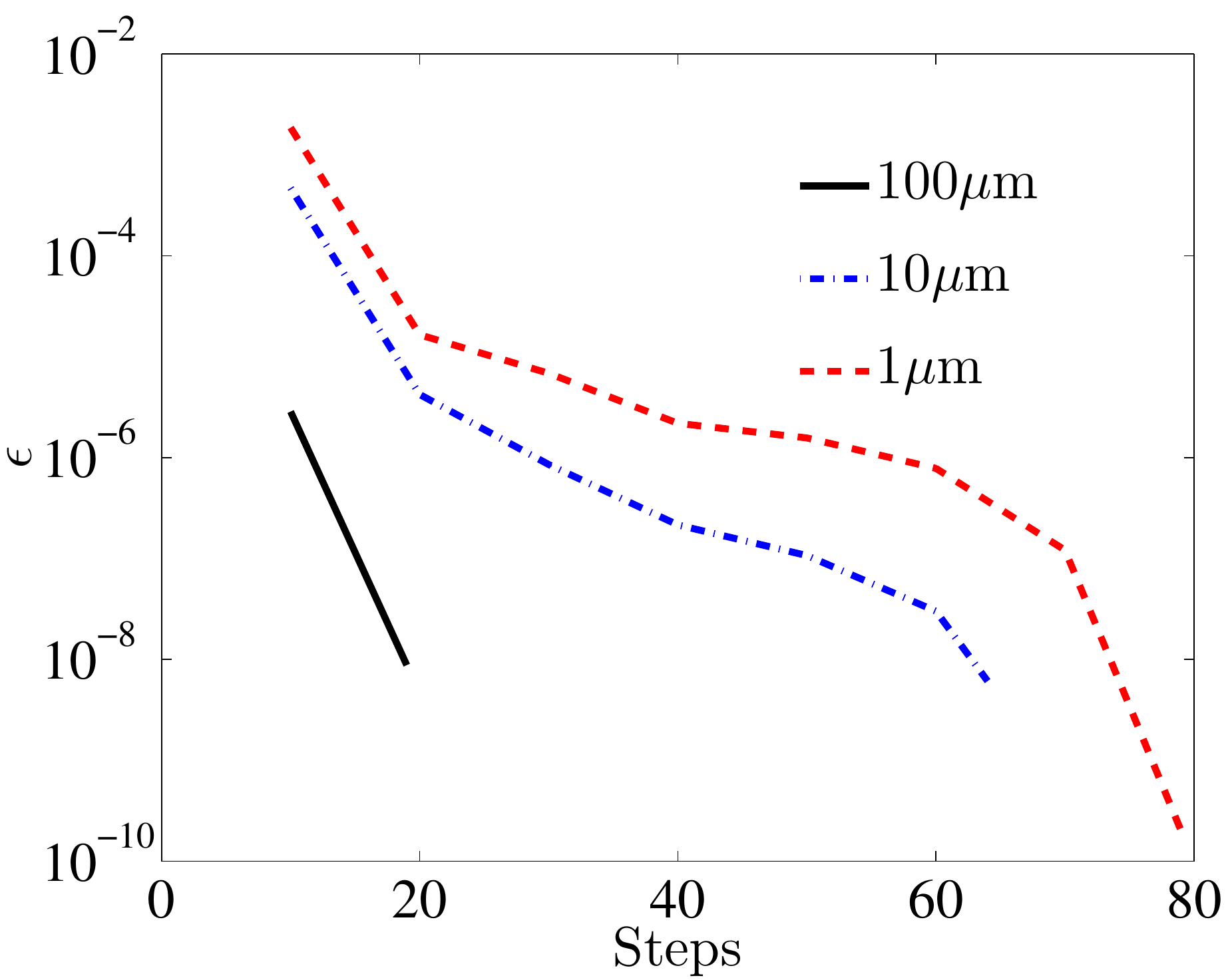}}~~
     \subfloat[]{\includegraphics[width=0.45\textwidth]{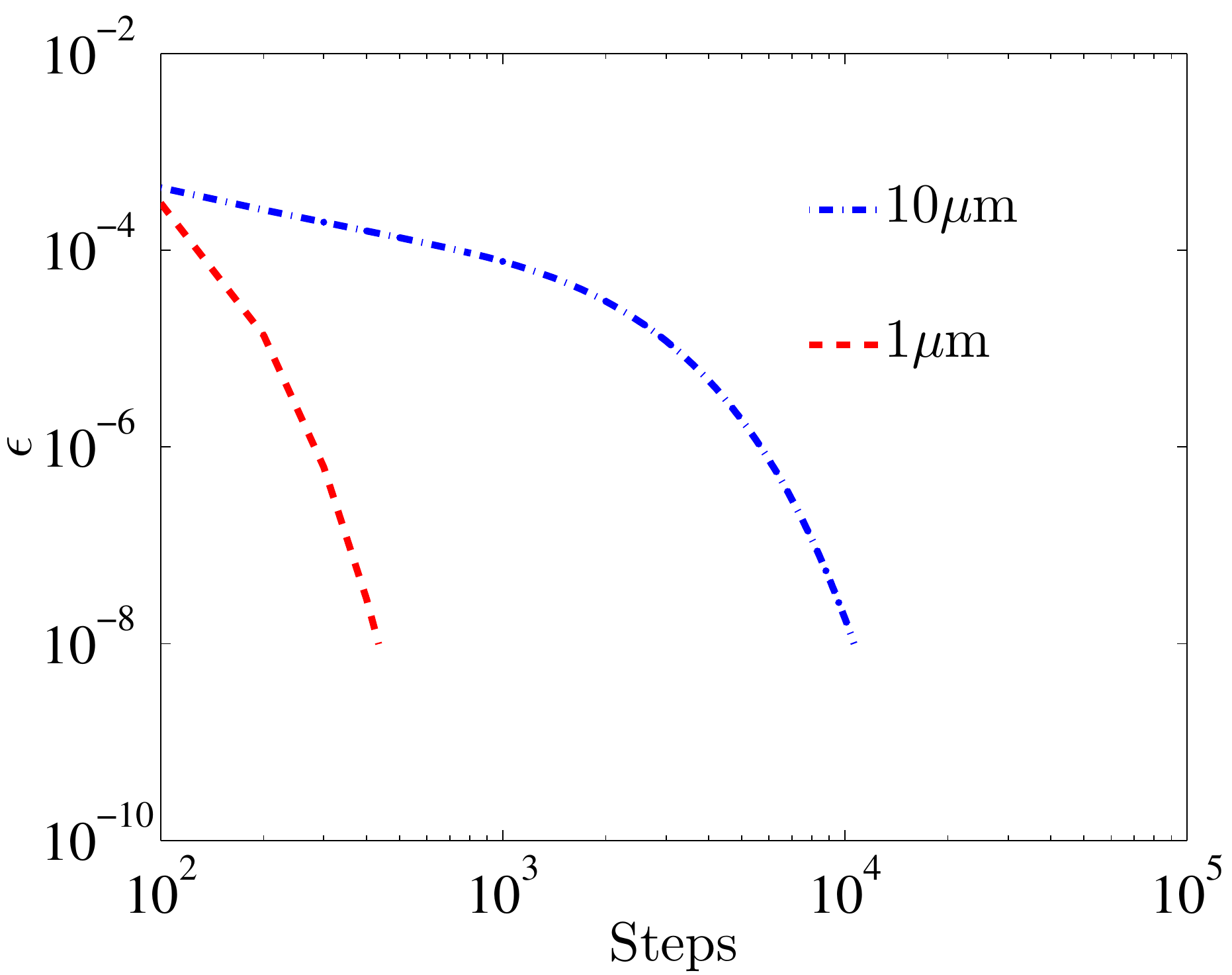}}~~\\
     \caption{Convergence history of the present scheme and the implicit DOM with different $L$ in the in-plane heat transfer. (a) Present scheme, (b) implicit DOM. }
     \label{historyinplane}
\end{figure}

\subsubsection{Isothermal solid wall heat transfer}

The thermalization boundary conditions are implemented on the left, right, top and bottom faces.
The temperature of the left face is fixed at $T_L=T_{\text{ref}} + \Delta T/2$, and the temperature of the other three faces is $T_L=T_{\text{ref}}-\Delta T/2$.
The front and back faces are set to be periodic.

The heat transfer with different $L$ is tested in this part.
The phonon dispersion is accounted with $N_B=20$.
As $L=100$nm, we set $N_x=N_y=50,~N_z=1$, $N_{\theta} \times N_{\varphi}=48 \times 24$ due to the highly non-equilibrium effects and MPI paralleling with 48 cores based on the solid angle space is used to save computation time.
For the other cases, the heat transfer comes close to that in the diffusive regime.
More cells ($N_x=N_y=100,~N_z=1$) have to be used, while the number of the discretized directions can reduce, for example $N_{\theta} \times N_{\varphi}=24 \times 24$.
The numerical results predicted by the present scheme are compared with \lei{those} obtained by the implicit DOM in~\cref{isothermal2D}.
Both the temperature and the heat flux predicted by the present scheme are in excellent agreement with those obtained by the implicit DOM.
\chuang{It can be found that the temperature jump happens on the left, top and bottom walls as $L=100$nm and $1\mu$m.
As $L$ increases from $100$nm to $10\mu$m, the non-equilibrium thermal effects decrease.}
The efficiency of the present scheme is tested and shown in Table.~\ref{2Defficiency}.
As $L=100$nm, both the implicit DOM and the present scheme reach convergence within $100$ iteration steps.
For the implicit DOM, as $L=1\mu$m, the convergence speed decreases much, and as $L=10\mu$m it is very hard to reach convergence.
But for the present scheme, convergence can be reached within $100$ steps for all cases.
As $L$ is larger than $1\mu$m, the present scheme is over ten times faster than the implicit DOM.

\begin{table}
\caption{The efficiency of the present scheme in isothermal solid wall heat transfer. Accelerate rate is the ratio of the CPU time between the implicit DOM and the present scheme. Steps mean the total iteration number. As $L=100$nm, MPI paralleling with 48 cores based on the solid angle space is used.}\vskip 0.2cm
\centering
\begin{tabular}{|*{6}{c|}}
 \hline
 \multirow{2}{*}{{\shortstack{$L$ }}}  & \multicolumn{2}{c|}{Present} & \multicolumn{2}{c|} {DOM} & \multirow{2}{*}{{\shortstack{Accelerate rate}}}\\
\cline{2-5}
 & Time (s) & Steps    & Time (s) & Steps   &  \\
   \hline
 10 $\mu$m & 196 & 72     & 52931 & 21840   & 270  \\
  \hline
 5 $\mu$m & 199 & 73     & 17476 & 7215   & 87.8  \\
\cline{1-3}
\cline{5-6}
 \hline
1 $\mu$m & 207 & 75  & 1917 & 785  & 9.3 \\
\cline{1-3}
\cline{5-6}
 \hline
  500 nm & 178 & 65  & 841 & 345 &   4.7 \\
 \hline
 100 nm & 416 & 65  & 399 & 66 &  0.96 \\
 \hline
\end{tabular}
\label{2Defficiency}
\end{table}

\begin{figure}[t]
     \centering
     \subfloat[]{\label{TB100nm}\includegraphics[width=0.33\textwidth]{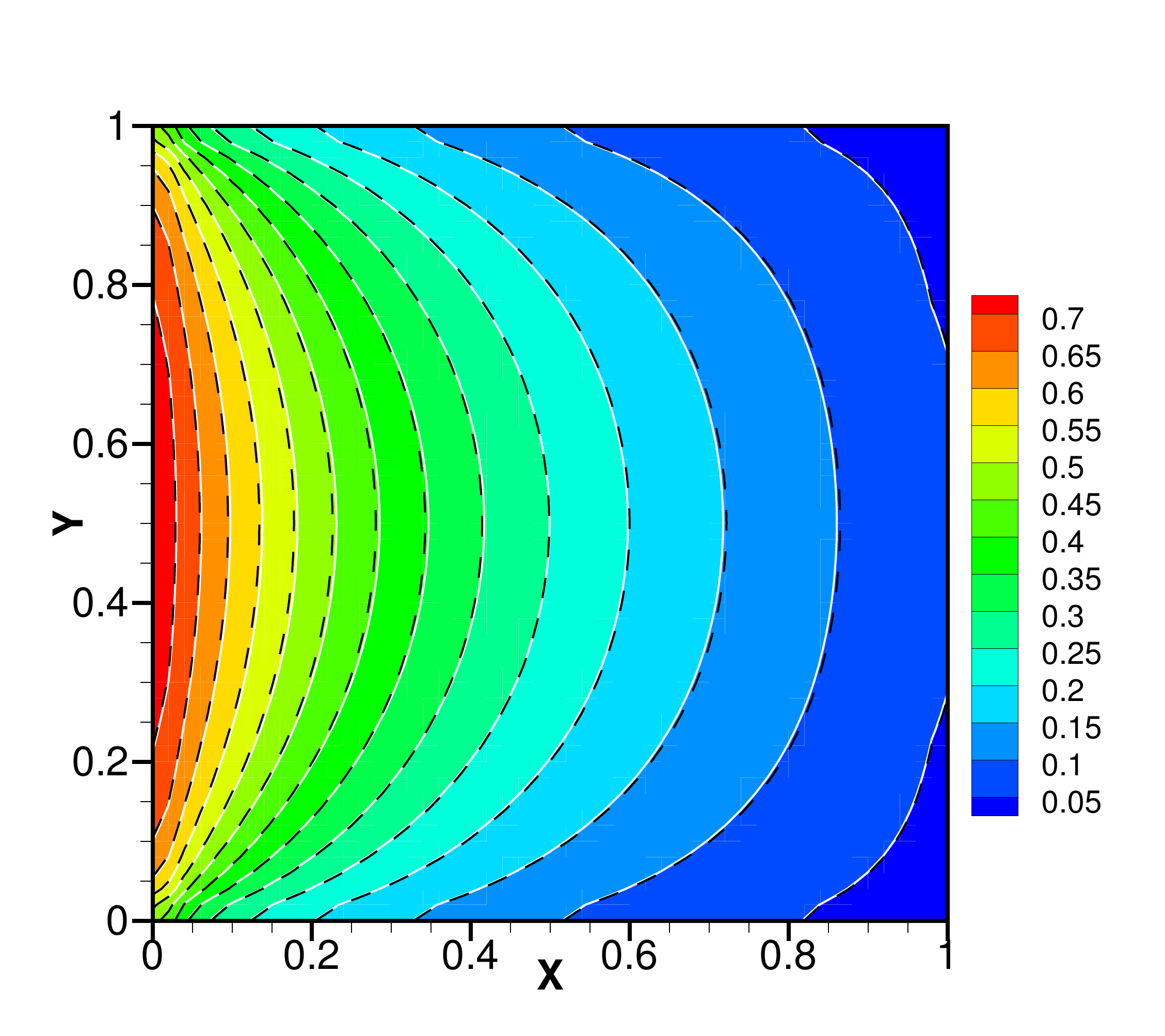}}~~
     \subfloat[]{\includegraphics[width=0.33\textwidth]{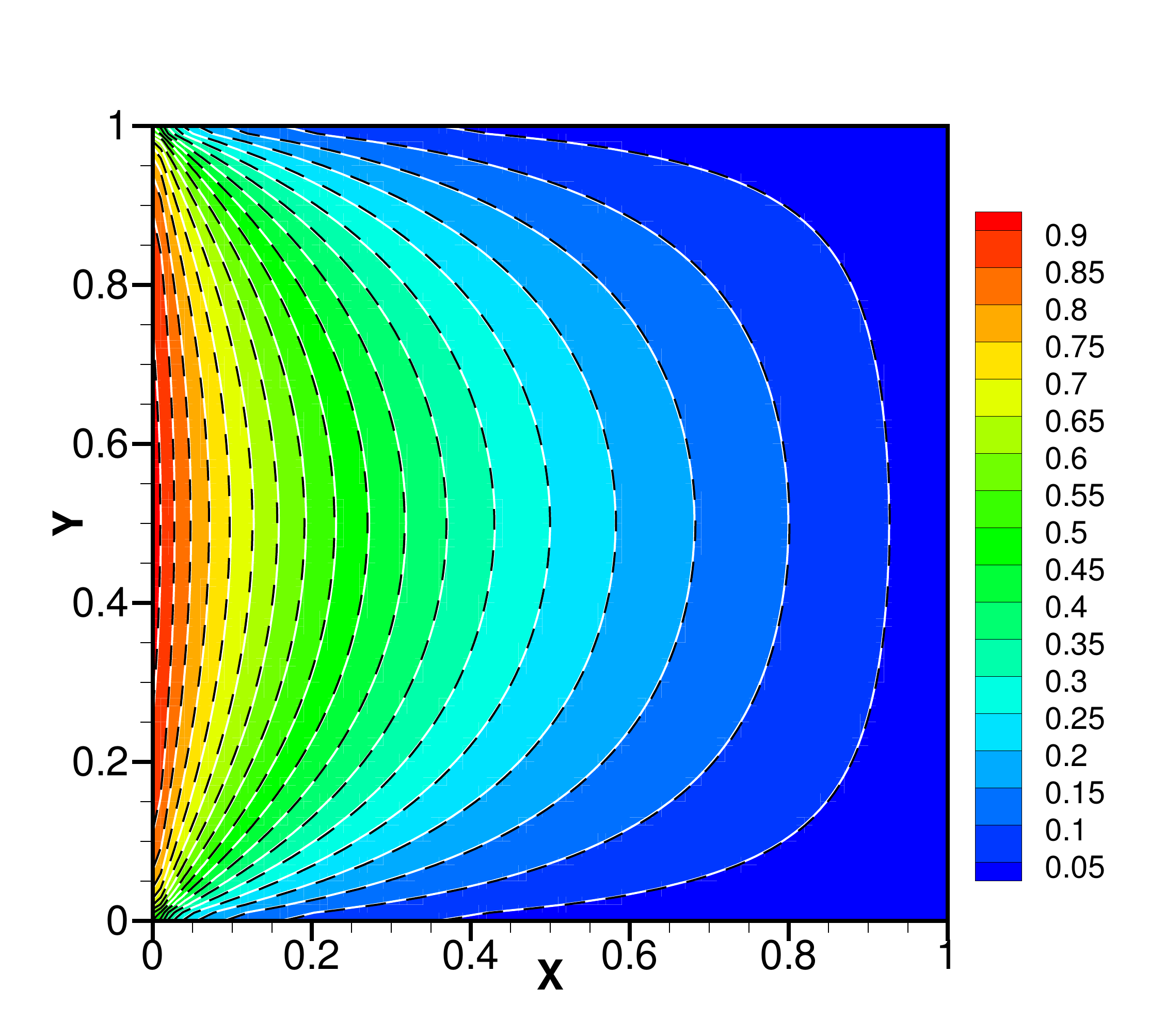}}~~
     \subfloat[]{\includegraphics[width=0.33\textwidth]{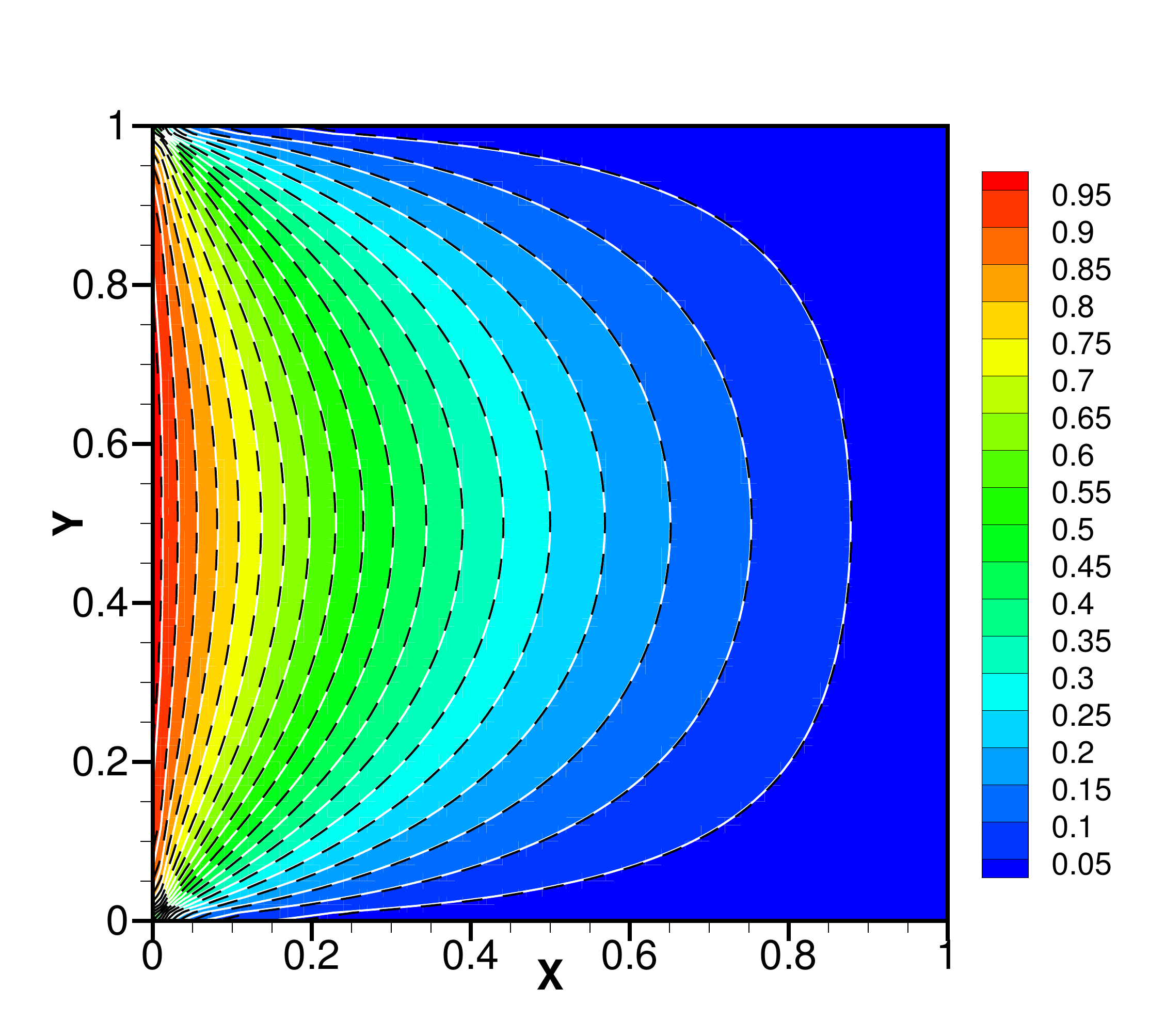}}~~\\
     \subfloat[]{\includegraphics[width=0.33\textwidth]{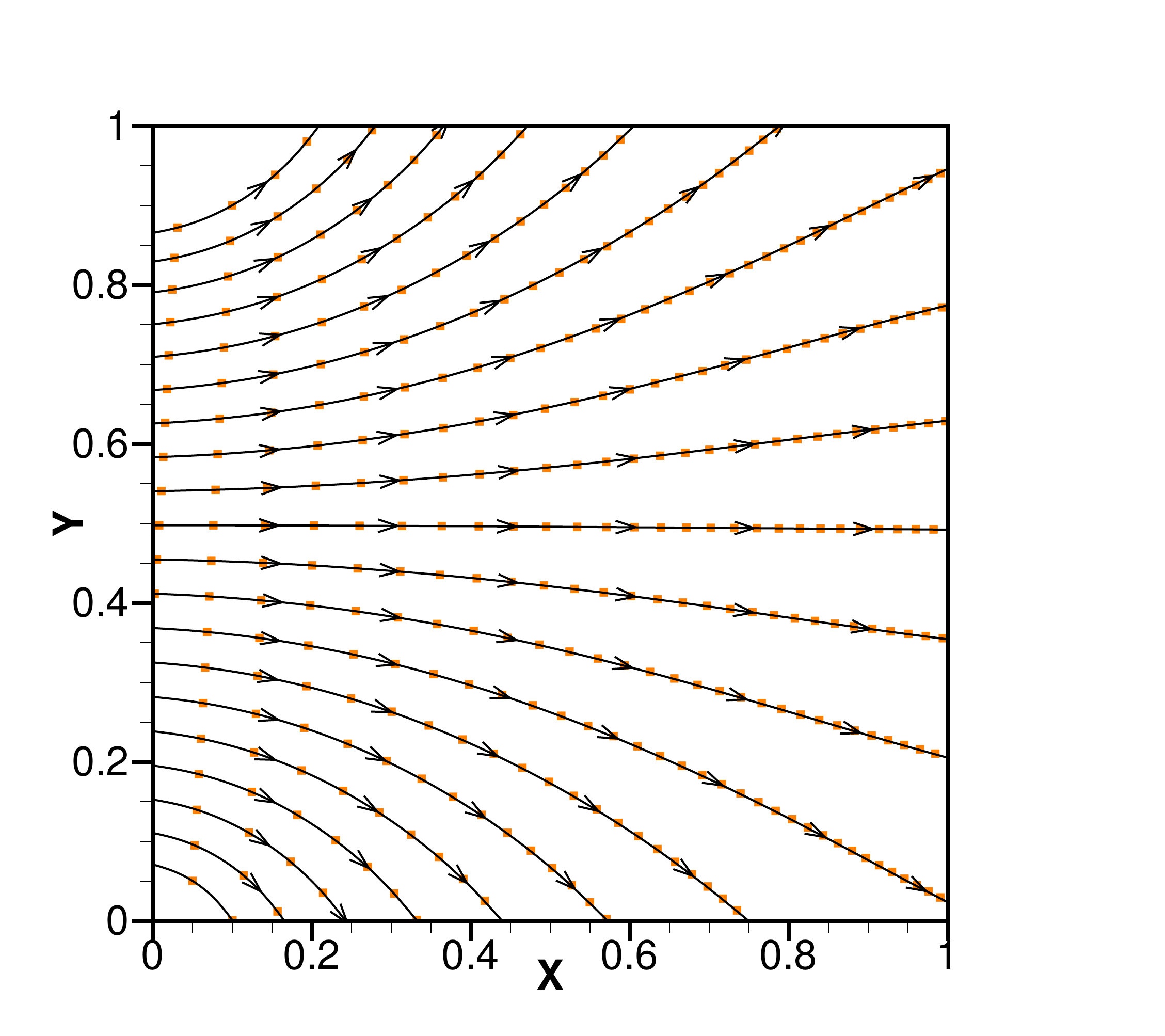}}~~
     \subfloat[]{\includegraphics[width=0.33\textwidth]{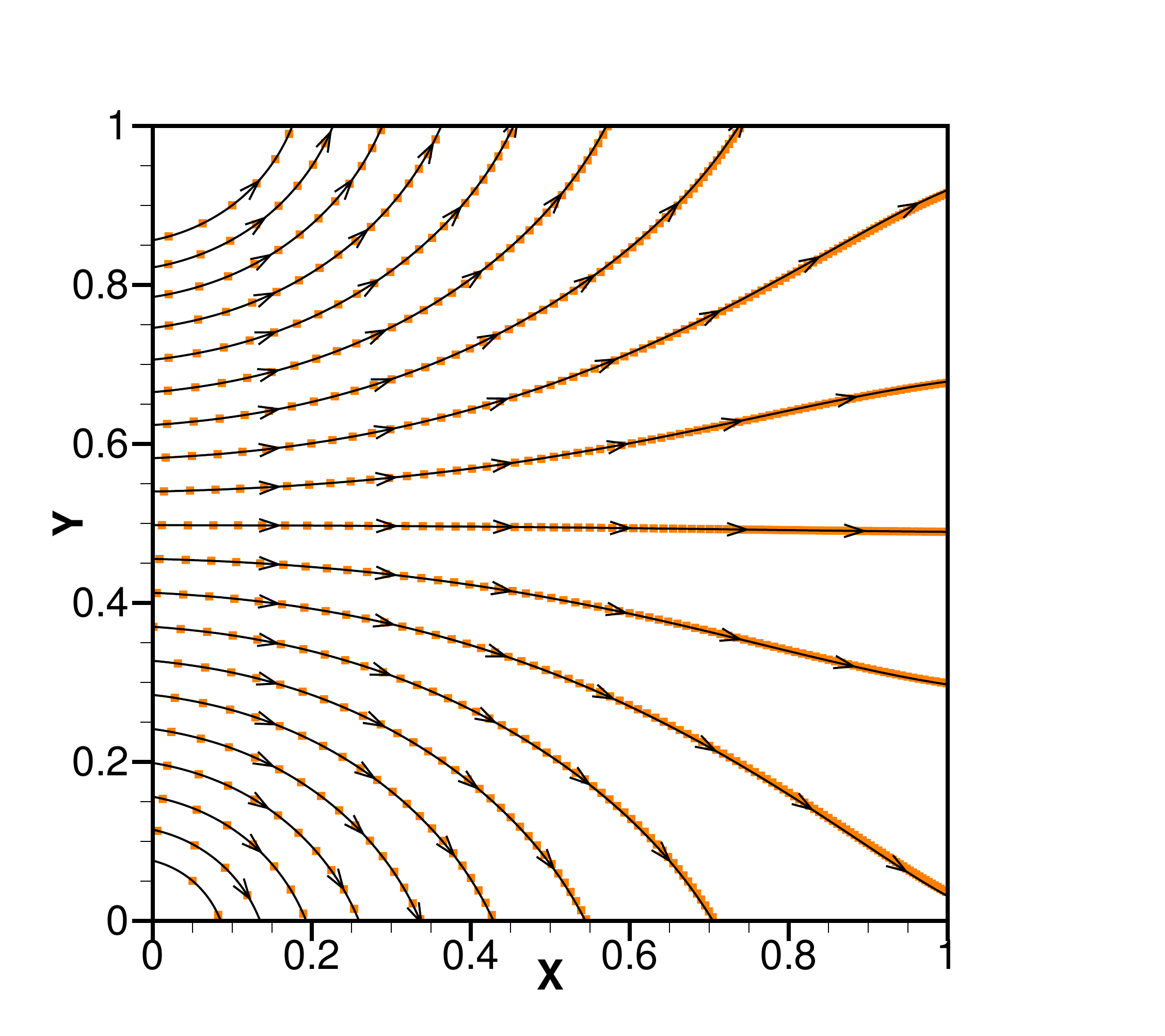}}~~
     \subfloat[]{\includegraphics[width=0.33\textwidth]{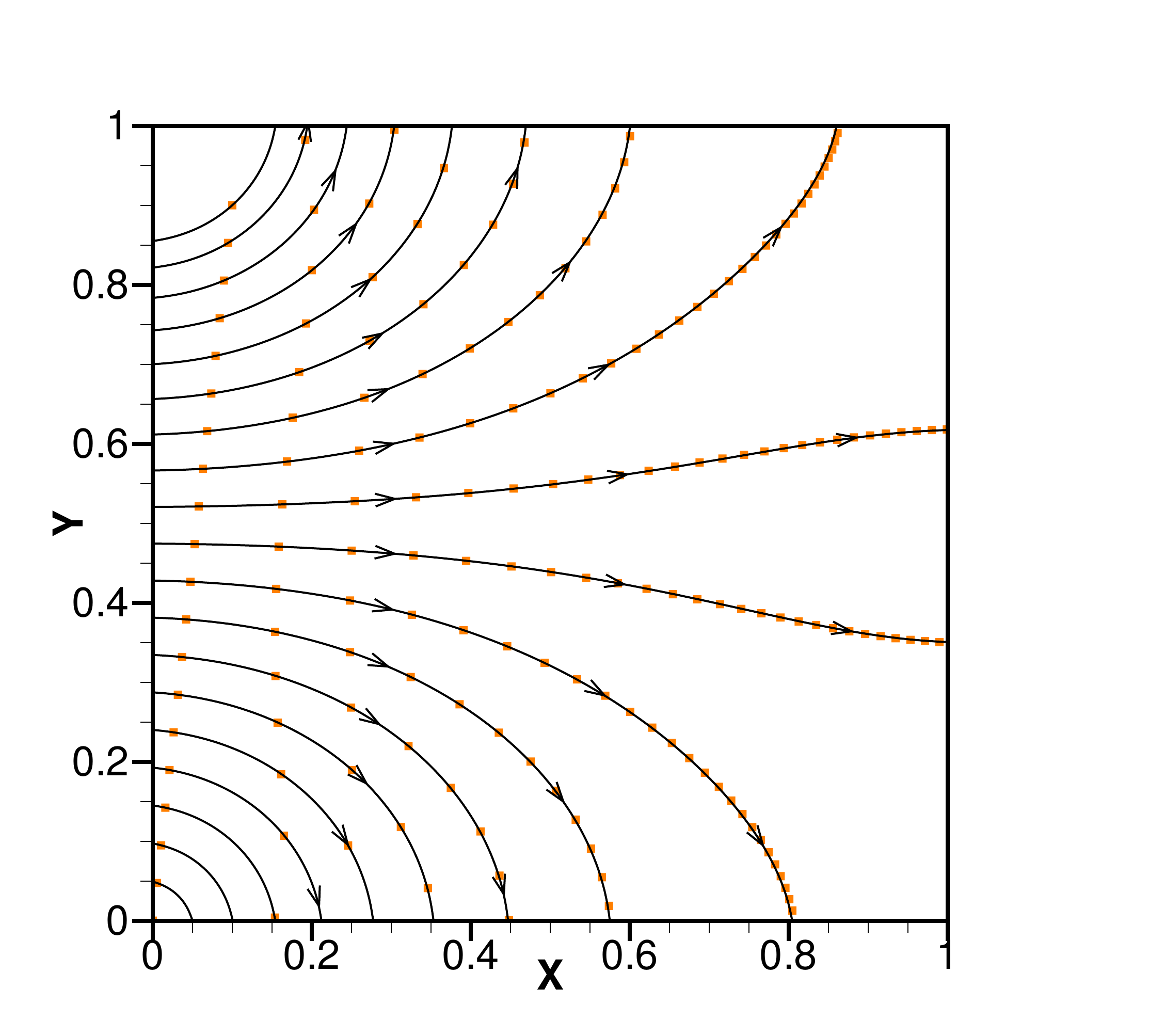}}~~\\
     \caption{Temperature contour and heat flux line of the isothermal solid wall heat transfer. $X$, $Y$ are the normalized coordinates, i.e., $X=x/L,~Y=y/L$. The normalized temperature is $(T-T_R)/{\Delta T}$. The above three figures are the temperature field and the bottom are the heat flux streamline. From the left to right are $L=100$nm, $1\mu$m, $10\mu$m, respectively. In the temperature contour, colored background with white solid line: implicit DOM; black dash line: present scheme. In the heat flux streamline, black line with arrowhead: implicit DOM; square orange dot: present scheme.  }
     \label{isothermal2D}
\end{figure}

\subsection{Three-dimensional  heat transfer}

Based on above subsections, it can be found that the present scheme converges much faster in the near-diffusive regime than the implicit DOM in 1D and 2D cases.
To better test the performances of the present scheme in the near-diffusive regime, a large-scale computation of a 3D device-like structure was undertaken~\cite{SyedAA14LargeScale}.
The geometry is shown in~\cref{3d_large}.
The length of the geometry in the $x$, $y$ and $z$ direction is $L_x,~L_y,~L_z$, respectively.
At the center of the top face ($z=0$), there is a square heating area with side length $L_h$.
The temperature of the heating area is $T_h =T_{\text{ref}}+\Delta T/2$.
At the bottom of the geometry ($z=L_z$), there is a cold area located at the center.
The side lengths are $L_c$ and $L_y$, respectively.
The temperature of the cold area is $T_c =T_{\text{ref}}-\Delta T/2$.
The other boundaries are all adiabatic.
The heat is generated at the top and dissipated at the bottom, which is like the thermal transport mechanism in a transistor.

In order to simulate this problem, the thermalization boundary conditions are implemented on the hot and cold area.
For the adiabatic boundaries, the diffusely reflecting boundary conditions are used.
We set $L_x=L_y=2L_z=4\mu$m, $L_h=L_c=1\mu$m,
$N_x \times N_y \times N_z =80 \times 80 \times 40 $, $N_B=20$ and $N_{\theta} \times N_{\varphi}=24 \times 24$, which is enough to capture the heat transfer process accurately.
Due to the large computational amount and memory requirement, we use the MPI paralleling computation with 192 cores based on the solid angle space.

The numerical results including the heat flux streamline and temperature contour are shown in~\cref{3D_xslice}.
From a global view (\cref{3dfig1}), the heat flux flows from the hot area to cold area and the temperature decreases gradually along the heat flux line.
From the temperature shown in~\cref{3dfig2,3dfig3}, it can be found that there is small temperature jump close to the hot area, which indicates the failure of the Fourier's law.
Furthermore, in this simulation, convergence is reached by $72$ steps.
Other simulations are also done with different numerical settings, as shown in Table.~\ref{3Defficiency}.
It can be observed that for all cases, convergence is reached within 100 steps in the near-diffusive regime.
In summary, the present scheme will be a powerful tool in simulating 3D large scale heat transfer, especially in the near-diffusive regime.

\begin{figure}[t]
 \centering
 \includegraphics[width=0.6\textwidth]{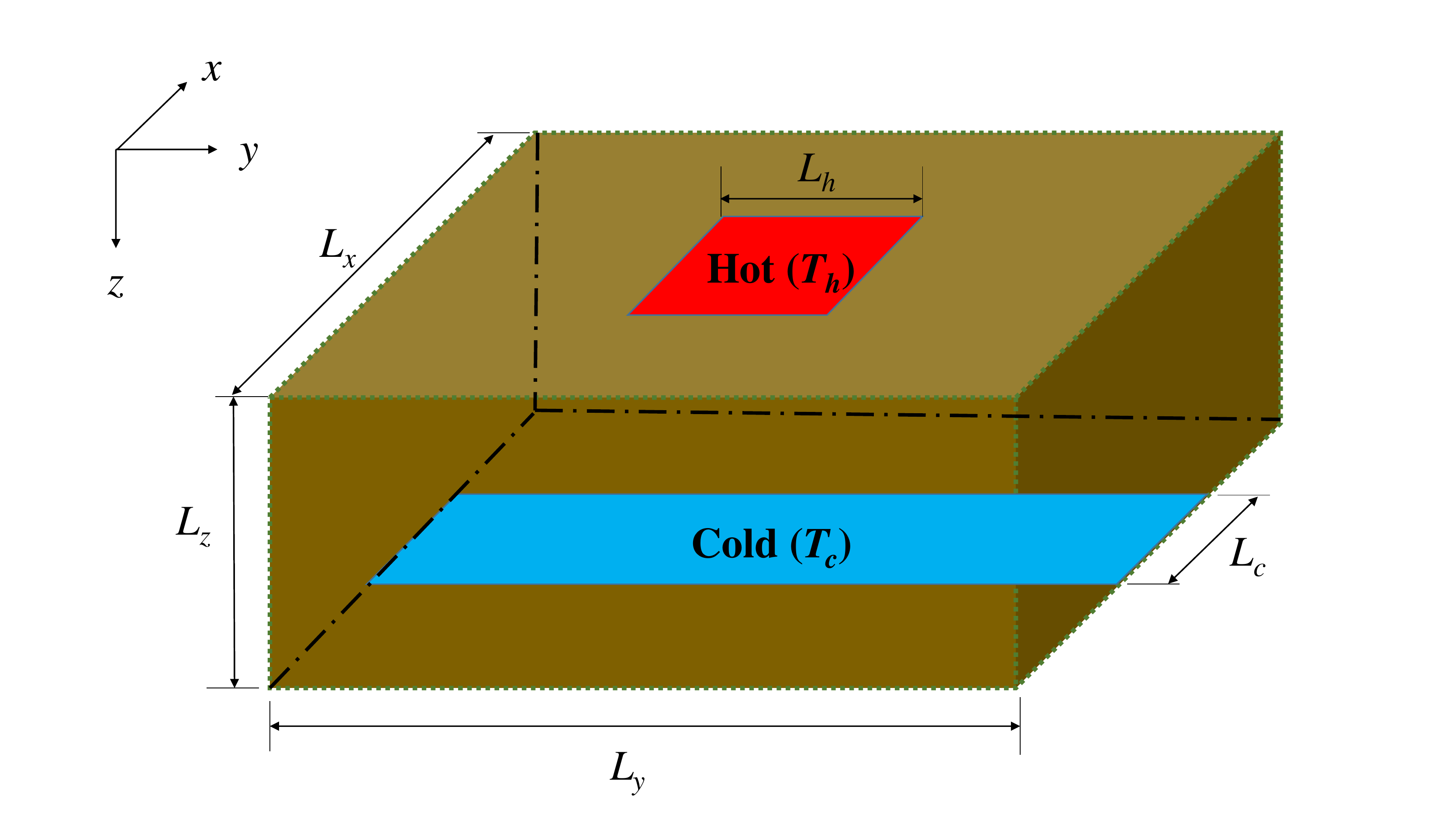}
 \caption{3D large scale heat transfer.}
 \label{3d_large}
\end{figure}

\begin{figure}[t]
     \centering
    \subfloat[]{\label{3dfig1}\includegraphics[scale=0.4,viewport=0 100 660 500,clip=true]{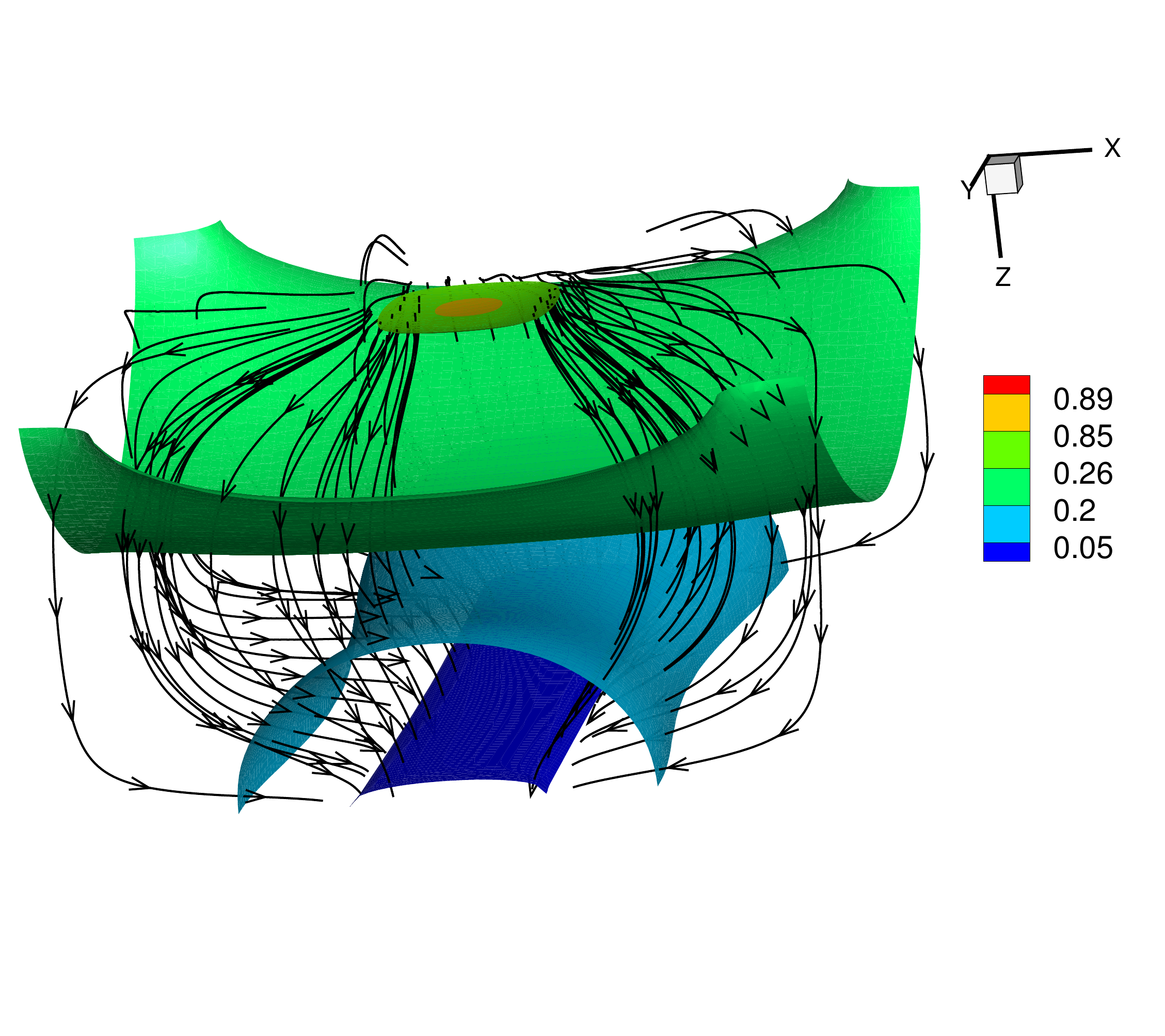}}\\
     \subfloat[]{\label{3dfig2}\includegraphics[scale=0.36,viewport=0 60 640 450,clip=true]{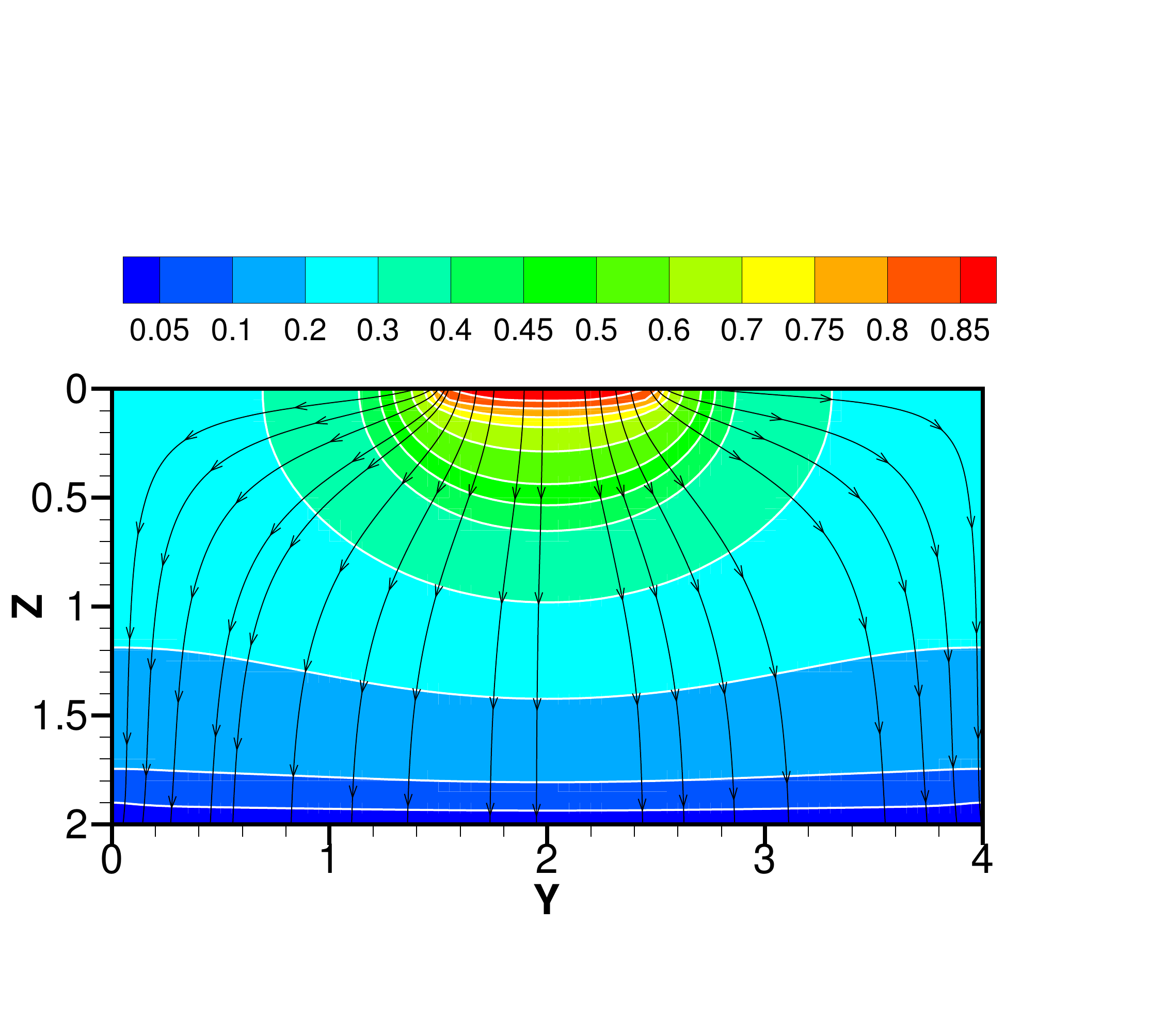}}~~
     \subfloat[]{\label{3dfig3}\includegraphics[scale=0.36,viewport=0 60 640 450,clip=true]{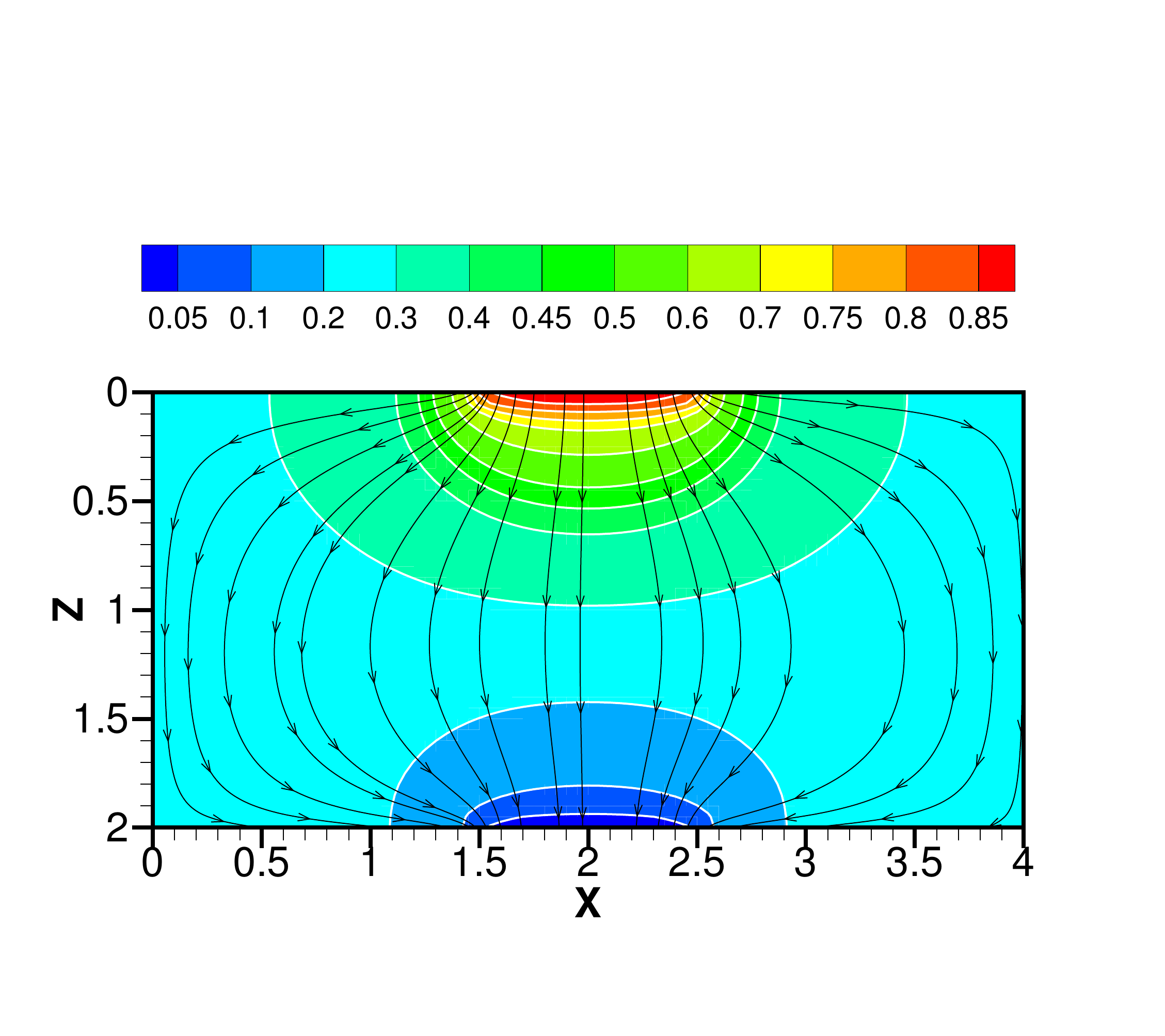}}~~\\
     \caption{Macroscopic distribution of the 3D heat transfer. Normalized temperature is $(T-T_c)/(T_h-T_c)$, and normalized coordinates are $X=x/L_0,~Y=y/L_0$, where $L_0=1\mu$m. (a) Temperature iso-surfaces and heat flux streamline,  (b) temperature contour and the heat flux streamline at $x=L_x/2$ slice, (c) temperature contour and the heat flux streamline at $y=L_y/2$ slice.  }
     \label{3D_xslice}
\end{figure}

\begin{table}
\caption{The efficiency of the present scheme in 3D heat transfer, where $N_B=20$, $L_h=L_c=L_x/4$.}\vskip 0.2cm
\centering
\begin{tabular}{|*{5}{c|}}
 \hline
Case & $L_x \times L_y \times L_z$ ($\mu$m$^3$) & $N_x \times N_y \times N_z$ &  $N_{\theta} \times N_{\varphi}$ & Steps  \\
\hline
 1  & $1 \times 1 \times 0.5  $ & $80\times 80 \times 40$ & $24 \times 24$ & 99 \\
%  \hline
% 2& $1 \times 1 \times 1  $ & $80\times 80 \times 80$ & $24 \times 24$ & 123 \\
  \hline
 2& $4 \times 4 \times 2  $ & $80\times 80 \times 40$ & $24 \times 24$ & 72 \\
 \hline
 3& $4 \times 4 \times 4  $ & $80\times 80 \times 80$ & $24 \times 24$ & 77 \\
  \hline
 4& $12 \times 12 \times 6  $ & $120\times 120 \times 60$ & $16 \times 16 $ & 67 \\
 \hline
\end{tabular}
\label{3Defficiency}
\end{table}

\section{CONCLUSIONS}\label{conclusion}

In this work, a synthetic iterative scheme is developed to accelerate convergence for the implicit discrete ordinate method in the near-diffusive regime based on the phonon Boltzmann transport equation.
\chuang{The key point of the present scheme is the introduction of the macroscopic synthetic diffusion equation for the temperature, which is exactly derived from the zero- and first-order moment equations of the phonon BTE and recovers the \lei{Fourier's heat conduction} equation correctly in the diffusive limit.
In the diffusion equation, the heat flux is separated into the Fourier part and the non-Fourier part.
The former contains the temperature diffusion information and the latter is obtained by the second-order moment of the distribution function, which captures the non-equilibrium phonon transport physics.
The phonon BTE and macroscopic diffusion equations are tightly coupled at two different levels.
At the macroscopic level, the diffusion equation provides the temperature
for the BTE;
at the \lei{mesoscopic} level, the BTE provides the second-order moment to the diffusion equation to describe the non-Fourier heat transfer.
%This synthetic iterative scheme strengthens the coupling of all phonons in the phase
%space to facilitate the fast convergence from the diusive to ballistic regimes
%The macroscopic diffusion equation separates the Fourier heat flux and the non-Fourier heat flux.
%In the diffusive limit, the
%The mathematical relations among the heat flux, temperature and the distribution function are built correctly by the phonon BTE and its moment equations.
%At the macroscopic level, the heat flux is separated into the Fourier part and the non-Fourier part in the first-order moment equation of the phonon BTE.
%Then the temperature can be updated by the non-Fourier heat flux by numerically enforcing the spatial divergence of the heat flux to be zero.
%At the microscopic level, the non-Fourier heat flux is obtained by the second-order moment of the distribution function, which is updated by the iterative solution of the phonon BTE with implicit DOM.
%In the diffusive regime, the non-Fourier heat flux tends to zero and the macroscopic equation recovers the Fourier law automatically.
%Therefore, the macroscopic equation converges very fast in the near-diffusive regime, which mitigates the weakness of the implicit DOM.
The efficient information exchange strengthens the coupling of all phonons in the phase space and makes the present synthetic scheme converge fast in the simulations of the steady heat transfer problems from diffusive to ballistic regimes.}

A number of numerical tests have confirmed that the present scheme can predict the thermal transport phenomena accurately in a wide range.
Furthermore, the present scheme accelerates convergence significantly in the near-diffusive regime, with about one to three orders of magnitude faster than the conventional implicit DOM.
For all cases considered in this study, including one-, two-, and three-dimensional problems, convergence can usually be reached within 100 steps in the near-diffusive regime.
%The present scheme will be an excellent and efficient tool to predict the heat transfer in those regimes that the traditional implicit DOM cannot work well.

We believe our method can be also used to construct fast convergence scheme for phonon hydrodynamic based on the Callaway's dual relaxation model~\cite{lee_hydrodynamic_2015,cepellotti_phonon_2015,wangmr17callaway}.

\section*{Acknowledgments}

This work was supported by the National Key Research and Development Plan (No. 2016YFB0600805) and the UK's Engineering and Physical Sciences Research Council (EPSRC) under grant EP/R041938/1.
%Chuang Zhang acknowledges the financial support of the National Science Foundation of China (11602091, 91530319).

\section*{References}

\bibliographystyle{elsarticle-num}

\bibliography{phonon}

\end{document}